\documentclass[prr,twocolumn,showpacs,amsmath,amssymb,superscriptaddress,floatfix,reprint, nofootinbib, longbibliography]{revtex4-1}
\usepackage{amsmath}
\usepackage{amssymb}
\usepackage{amsfonts}
\usepackage{framed}
\usepackage{listings}
\usepackage{enumerate}
\usepackage{latexsym}
\usepackage{xcolor}

\usepackage{amsthm}
\theoremstyle{definition}

\usepackage{hyperref}

\usepackage{tensor}

\usepackage{psfrag}
\usepackage{xcolor}
\usepackage{float}
\usepackage{soul}

\usepackage{graphicx}
\usepackage{subfigure}
\usepackage{tikz-cd}

\newcommand{\beq}{\begin{equation}}
\newcommand{\eeq}{\end{equation}}
 
\newcommand{\UU}{\mathrm{U}}

\newcommand{\eqnref}[1]{Eq.~(\ref{#1})}

\newcommand{\Epar}{\mathbf{E}_\parallel}

\newcommand{\Eperp}{\mathbf{E}_\perp}

\newcommand{\ket}[1]{|#1\rangle}

\begin{document}

\title{Quantum Many-Body Topology of Quasicrystals
}

\author{Dominic V. Else}
\affiliation{Department of Physics, Massachusetts Institute of Technology, Cambridge, MA 02139, USA}
\author{Sheng-Jie Huang}
\affiliation{Condensed Matter Theory Center and Joint Quantum Institute, Department of Physics, University of Maryland, College Park, Maryland 20742-4111, USA}
\author{Abhinav Prem}
\affiliation{Princeton Center for Theoretical Science, Princeton University, Princeton, NJ 08544, USA}
\author{Andrey Gromov}
\affiliation{Brown Theoretical Physics Center and Department of Physics, Brown University, RI 02912, USA}

\begin{abstract}
In this paper, we characterize quasicrystalline interacting topological phases of matter i.e., phases protected by some quasicrystalline structure.
We show that the elasticity theory of quasicrystals, which accounts for both ``phonon" and ``phason" modes, admits non-trivial quantized topological terms with far richer structure than their crystalline counterparts. 
We show that these terms correspond to distinct phases of matter and also uncover intrinsically quasicrystalline phases, which have no crystalline analogues. For quasicrystals with internal $\UU(1)$ symmetry, we discuss a number of interpretations and physical implications of the topological terms, including constraints on the mobility of dislocations in $d=2$ quasicrystals and a quasicrystalline generalization of the Lieb-Schultz-Mattis-Oshikawa-Hastings theorem. We then extend these ideas much further and address the complete classification of quasicrystalline topological phases, including systems with point-group symmetry as well as non-invertible phases. We hence obtain the ``Quasicrystalline Equivalence Principle," which generalizes the classification of crystalline topological phases to the quasicrystalline setting.
\end{abstract}

\date{\today}
\maketitle
%


\section{Introduction}
\label{sec:intro}

Spurred by the theoretical prediction~\cite{kanemele,bernevigzhang,fukanemele,moorebalents,roy2009} and experimental discovery~\cite{konig2007,hsieh2008} of topological band insulators, there has been remarkable progress in the topological classification of gapped quantum matter~\cite{hasankanermp,qizhangrmp}. By now, it is well-established that topological insulators and other free-fermion topological states~\cite{kitaev2009,ryu2010} belong to the much larger family of symmetry protected topological (SPT) phases, which encompass strongly correlated systems with internal (on-site) symmetries~\cite{guwen2009,pollmann2010,fidkowski2011,chen2011,schuch2011,levingu2012,chen2013,guwen2014}. SPT phases lack spontaneous symmetry breaking and are instead characterised by a bulk gap to all excitations, a unique ground state (with periodic boundary conditions), and non-trivial surface states which are robust against arbitrary local symmetry-preserving perturbations~\cite{colemanreview,senthilSPTrev}. The ground state of a non-trivial SPT can be adiabatically
connected to a trivial product wave function only if the bulk gap is closed or if the symmetries protecting the system are explicitly broken. Thus, two states which are smoothly connected absent any symmetries may belong to distinct SPT phases once a symmetry is enforced. 

The topological classification has been further extended to include crystalline point group and space group symmetries, such as spatial reflection or rotation, which are of intrinsic interest, being crucial for understanding phenomena in solids. Non-interacting fermionic systems protected by such symmetries, also called topological crystalline materials (see Refs.~\cite{andofu2015,chiu2016rmp} and references therein), are particularly well-understood and include so-called higher-order topological insulators (HOTIs)~\cite{neupert2018rev}. As a consequence of crystalline symmetry, a $d$-dimensional higher-order topological phase is gapped everywhere except on a $(d-n)$-dimensional surface (with $n>1$), such that a non-trivial 3d HOTI can host gapless hinge ($n=2$) or corner ($n=3$) modes. More recently, a general framework for interacting topological phases protected by crystalline symmetries---crystalline SPT (cSPT) phases---has emerged and is conjectured to give a complete classification of crystalline topological matter~\cite{song2017,buildingblock,Thorngren_1612,Else_1810}. 

Given the progress in classifying crystalline topological phases, a natural question is whether topological phenomena can be protected by the structure of \emph{quasicrystals}, which possess long-range orientational order but are non-periodic~\cite{shechtman1984,levine1984}, and hence occupy a regime intermediate between periodic and amorphous structures. The long-range order of quasicrystals manifests in sharp Bragg peaks, and although lacking translation symmetry, quasicrystals can still have a notion of rotation symmetries (often crystallographically forbidden). Moreover, quasicrystals can always be understood as an incommensurate projection of a higher-dimensional periodic lattice to the physical space. This latter fact has been exploited for understanding topological phases of quasicrystals, whose properties can sometimes be inferred from those of a higher dimensional periodic system. For instance, certain 1d quasiperiodic systems have been shown~\cite{kraus2} to inherit the topological indices and edge states of the ubiquitous 2d Harper-Hofstadter model~\cite{harper1955,hofstadter1976}. Indeed, a plethora of intriguing phenomena have been unearthed in the context of non-interacting quasicrystalline topological phases~\cite{lang2012,kraus1,kraus2,kraus3,dana2014,singh2015,tran2015,prodan2015,loring2016,bandres2016,quasicrystalQSH,spinbott,zhou2019,lin2020simulating}, including HOTI phases without crystalline analogues~\cite{varjas2019,chen2020,spurrier2020}.

Besides its intrinsic theoretical interest, the study of quasicrystalline topological matter has immediate experimental relevance, especially given the possibility of creating and manipulating synthetic lattices with quasiperiodic structures on various platforms, including ultra-cold atoms and optical cavities, amongst others~\cite{SanchezPalencia_0502,schreiber2015,verbin2015,vignolo2016,baboux2017,girovsky,collins2017,drost,dareau2017,viebahn2019,Gautier_2010}. A burgeoning platform for realising electronic quasicrystals is twisted bilayer graphene~\cite{ahn2018,yao2018,tbgqc}, which provides an exciting avenue for exploring the interplay between quasicrystalline order, topology, and strong interactions. However, there has been little progress towards the classification and characterisation of strongly correlated quasicrystalline topological phases. While some studies have considered the stability of topologically non-trivial quasiperiodic spin chains to interactions~\cite{lado2019,liu2020,prodan2020}, a general classification akin to that of cSPTs has remained far from complete. In this paper, we fill this gap by proposing a general classification of quasicrystalline topological phases.

We proceed by first characterising cSPT phases in terms of their response to elastic deformations, as captured by the presence of a quantized topological response term in the action. Espousing this \textit{topological elasticity theory} perspective proves particularly efficacious when generalising to quasicrystals, whose elasticity theory is distinguished by the presence of both ``phonon" and ``phason" modes~\cite{lubensky1985,levine1985,lubensky1986,lubensky1986b,ding1993}. In particular, we show that \textit{despite} the absence of any translation symmetry, the elasticity theory of quasicrystals admits quantized topological terms with a far richer structure than their crystalline counterparts. 

For $d$-dimensional quasicrystalline SPTs with internal $\UU(1)$ symmetry, we enumerate several allowed topological terms. These terms, which we show correspond to distinct quasicrystalline phases of matter, include terms describing the quantized response to a background $\UU(1)$ gauge field; topological theta terms; and Wess-Zumino type topological terms. The latter two classes, which do not require $\UU(1)$ symmetry, are intrinsically quasicrystalline i.e., have no crystalline analogues. We discuss several interpretations and consequences of the term describing the response to a background $\UU(1)$ gauge field, including the mobility restrictions it imposes on dislocations in a $d=2$ quasicrystal. We also show that our results reproduce known results in the classical limit and in the limit of non-interacting fermions. However, an important aspect of our approach is that it naturally incorporates the role of quantum fluctuations and strong correlations, and hence also accounts for intrinsically interacting topological phases i.e., those with no non-interacting analogue.

We then extend these ideas much further and address the complete classification of quasicrystalline topological phases. As a first non-trivial extension, we consider quasicrystalline SPTs with only internal symmetries and no point-group symmetry. We show that, despite the absence of translation symmetry, such phases are partially classified by generalizations of the ``weak invariants"~\cite{weakSPT2016,cheng2019} that appear in the classification of SPT phases with translation symmetry. We also find additional \textit{intrinsically} quasicrystalline topological phases, which have no crystalline analogue. Finally, we provide a general classification which accounts for point-group symmetry and also extends to symmetry enriched topological phases; this is encapsulated in the \textbf{Quasicrystalline Equivalence Principle}, which extends the classification of crystalline topological states to the quasicrystalline setting and represents our main result.

The rest of this paper is organised as follows: in Sec.~\ref{sec:summary}, we give a pedagogical overview of our approach and summarise the main results. We discuss the topological elasticity theory for cSPTs with lattice translation and $\UU(1)$ symmetry in Sec.~\ref{sec:crystal_elasticity}. Sec.~\ref{sec:qcreview} reviews the primary concepts regarding quasicrystals that we will need throughout this paper. This is where define what we mean by a quasicrystalline topological phase and introduce the phase angle fields that parameterize elastic deformations in a quasicrystal; this differs from the usual description in terms of ``phonon" and ``phason" variables, so even those familiar with the elasticity theory of quasicrystals are encouraged not to skip this section. In Sec.~\ref{sec:qc_topoterm}, we generalize the topological elasticity theory to quasicrystals and show that the topological terms correspond to distinct quasicrystalline phases of matter; their classification in terms of integral cohomology is also stated here. Secs.~(\ref{sec:mobility}--\ref{sec:lsmoh}) are devoted to physical implications and interpretations of the topological terms for quasicrystals with $\UU(1)$ symmetry: in Sec.~\ref{sec:mobility}, we show how the topological term constrains the mobility of dislocations in 2d quasicrystals; in Sec.~\ref{sec:lsmoh} we formulate a quasicrystalline version of the LSMOH theorem; and in Sec.~\ref{sec:interpret}, we discuss some simple interpretations of the topological invariants. Finally, we present the general classification of quasicrystalline topological phases in Sec.~\ref{sec:general}, which includes systems with and without point-group symmetry as well as non-invertible phases. We conclude with a discussion of open questions and future directions in Sec.~\ref{sec:cncls}.  


\section{Summary of approach and main results}
\label{sec:summary}

\subsection{Definition of quasicrystalline topological phases}
\label{subsec:defn_1}

The usual definition of an SPT or symmetry-enriched topological (SET) phase is that of a family of gapped Hamiltonians that can be continuously deformed into each other while respecting the relevant symmetries, and without closing the bulk gap~\cite{Chen_1004}. If the symmetries include crystalline symmetries, then this constitutes a crystalline SPT or SET phase. The main point of this paper is to generalize these ideas to quasicrystals. However, in this case the definitions become somewhat more subtle, because unlike crystals, quasicrystals do not have any exact lattice translation symmetry, and while point-group symmetry can be defined for a quasicrystal, it is not literally a symmetry of the Hamiltonian \cite{Mermin_1992}. Instead, we will consider families of gapped Hamiltonians that can be continuously deformed into each other while preserving some notion of quasicrystallinity. We will give the precise definition later on (see Sections \ref{subsec:defn_2} and \ref{sec:general}).

Finally, let us note that a real quasicrystal, as in the crystalline case, will always possess gapless phonon modes due to the spontaneously broken continuous translation symmetry. Therefore, as in the case of crystalline topological phases, when we talk about gapped Hamiltonians, what we really mean is that there exists a gap for the remaining degrees of freedom after the phonons (and in the case of quasicrystals, the phasons -- see below) have been gapped out by adding a pinning term to the Hamiltonian that explicitly breaks the continuous translation symmetry (for example, by applying a periodic or quasiperiodic potential to pin the atoms to particular locations in space). It is in this case that the classification of topological phases can be made totally precise. However, much of the physics we discuss will still be relevant for the real crystal/quasicrystal where the phonons and phasons are gapless; we will return to this point later on.

\subsection{Understanding crystalline topological phases through elasticity theory}
\label{subsec:elasticity}

In general, symmetry-protected topological phases of matter can be characterized by their response to background gauge fields; for example, quantum Hall states are characterized by their Hall conductance. Attempts have been made~\cite{Thorngren_1612,Manjunath_2005,manjunath2021crystalline,gromov2019chiral} to generalize this to ``crystalline gauge fields,'' which are gauge fields for crystalline symmetries; however, the results are rather formal and difficult to interpret, and certainly it is not clear how to generalize them to quasicrystalline systems, which do not have exact crystalline symmetries to ``gauge.''

A core idea of this paper is that we should instead characterize crystalline topological phases through their response to elastic deformations (i.e., to phonons), which is much more physically meaningful in any case. This immediately makes it clear how to generalize to quasicrystalline topological phases; one simply needs to take into account that the elasticity theory of quasicrystals is richer than that of crystals, since in addition to phonon modes, they also host phason modes.

To illustrate the idea, let us recall that ``integer'' or ``invertible" topological phases (i.e. those without topologically non-trivial excitations) in a $d$-dimensional crystal with $d$ lattice translation symmetries and a $\UU(1)$ charge conservation symmetry have an integer invariant $\nu$, representing the charge per unit cell. For a system with invariant $\nu$, we claim that the topological response of such a system to elastic deformations and to a background $\mathrm{U}(1)$ gauge field $A_\mu$ is characterized by a quantized topological term that appears in the action (previous works discussing quantized topological terms in elasticity theory include Refs.~\cite{Cvetkovic_0508,Nissinen_1803,Nissinen_1812,Nissinen_2008,Nissinen_2009}). For example, in $d=2$, this term takes the form
\begin{equation}
\label{eq:cryst_topterm}
    \int \frac{\nu}{8\pi^2} \epsilon^{\mu \nu \lambda} \epsilon_{IJ} A_\mu \partial_\nu \theta^I \partial_\lambda \theta^J d^2 x dt,
\end{equation}
where the phase angle fields $\theta^1(\mathbf{x},t)$ and $\theta^2(\mathbf{x},t)$ are related to the phonon modes and are defined below. Furthermore, we will argue that, once the phason modess are included, these phase angles get generalized, for a 2-dimensional system, to $D$ phase angle fields $\theta^1, \cdots, \theta^D$ for some $D > 2$. Then one can generalize the above topological term to\footnote{Throughout this paper, Greek indices $\mu,\nu,\lambda$ take values $0,\cdots,d$; lower-case Latin indices $i,j,k$ run from $1,\cdots,d$; and upper-case Latin indices run from $1,\cdots,D$. Recall that for crystals, $D=d$.}
\begin{equation}
    \int \frac{1}{8\pi^2} C_{IJ} \epsilon^{\mu \nu \lambda} A_\mu \partial_\nu \theta^I \partial_\lambda \theta^J d^2 x dt,
\end{equation}
where $C^{IJ}$ can be any anti-symmetric $D$-dimensional integer matrix. Thus, such matrices give a partial classification of quasicrystalline topological phases with $\UU(1)$ symmetry. In general space dimension $d$, the analogous partial classification is by $D$-dimensional rank-$d$ integer tensors that are anti-symmetric in each pair of indices. Another way to say this is that there is an independent integer-valued invariant for each size-$d$ subset of the set $\{ 1, \cdots, D \}$.

In this paper, we will discuss various interpretations and implications of these invariants, such as:
\begin{itemize}
\item We will show that these invariants determine which directions a dislocation in a quasicrystalline topological phase can move in. This generalizes the familiar observation that in a crystal, dislocations can move only in the direction of their Burgers vector.
\item Since these invariants determine the average charge density as in the crystalline case, they allow one to formulate the quasicrystalline version of the Lieb-Schultz-Mattis-Oshikawa-Hastings (LSMOH) theorem \cite{Lieb_1961,Oshikawa_9610,Oshikawa_9911,Hastings_0411}; recall that the crystalline version of this theorem states that for an integer topological phase, the charge per unit cell must be an integer. The quasicrystalline version similarly specifies which average charge densities are permissible.
\item We give a generalization of the interpretation of $\nu$ as the charge per unit cell in the crystalline case; we argue that in a construction of a quasicrystal by tiling space through a set of tiles, the integer invariants $C^{IJ}$ can be identified as the charge bound to each kind of tile.
\end{itemize}

We wish to emphasize that, although when studying topological phases one traditionally considers only the electronic sector, the quasicrystalline invariants we discuss can also be applied to the atomic lattice of a quasicrystalline solid, where we identify the $\mathrm{U}(1)$ symmetry as representing conservation of number of atoms (one can of course have a generalization in which there is more than one conserved species of atoms). The invariants then coincide with ``stoichiometric coefficients for quasicrystalline compounds'' that have long been studied in connection with quasicrystals \cite{Kalugin_1989,numbertheory92}. However, while such quantities were previously studied assuming that the positions of the atoms are classical points, our new perspective makes it clear that the invariants are still well-defined and quantized in the presence of quantum fluctuations of the atomic locations.

\subsection{The general classification}

In this paper, we also go further and extend these ideas to address the complete classification of quasicrystalline topological phases. One can distinguish between quasicrystals with non-trivial point group symmetry and those without any point-group symmetry. 

\subsubsection{Invertible topological phases without point-group symmetry}
\label{subsubsec:weak}

In the case without point-group symmetry, we show that ``integer'' i.e., invertible, topological phases with internal symmetry $G$ in a quasicrystalline system with $D$ independent elastic modes (phonons and phasons) are classified by 
\begin{equation}
\label{eq:weak_invariants}
\bigoplus_{k=0}^{D} \mathcal{Q}_{d-k}^{\times \begin{pmatrix} D \\ k \end{pmatrix}} .
\end{equation}
Here $\mathcal{Q}_p$, for $p \geq 0$, is the classification of invertible topological phases with internal symmetry $G$ in $p$ spatial dimensions. Observe that in the crystalline case ($D = d$), \eqnref{eq:weak_invariants} reduces to the familiar classification of so-called ``weak invariants.'' The invariants discussed in Section \ref{subsec:elasticity} above correspond to the $k=d$ term in \eqnref{eq:weak_invariants} when the internal symmetry is $G = \UU(1)$ (recall that the classification of $0$-dimensional topological phases with $\mathrm{U}(1)$ symmetry is $\mathcal{Q}_0 = \mathbb{Z}$ i.e., the $\mathrm{U}(1)$ charge, hence why those invariants are integer-valued).

However, in the quasicrystalline case there is an additional phenomenon that has no crystalline analog. For $D > d$, $d-k$ can become negative in the sum \eqnref{eq:weak_invariants}. Topological phases in negative spatial dimension might not seem to make much physical sense, but it turns out that to get the correct classification of quasicrystalline topological phases it is necessary to define
\begin{equation}
\mathcal{Q}_{-1} = \begin{cases} \mathbb{Z}_2 & \mbox{$G$ contains time-reversal} \\
0 & \mbox{otherwise},
\end{cases}
\end{equation}
as well as
\begin{equation}
\mathcal{Q}_{-2} = \begin{cases} 0 & \mbox{$G$ contains time-reversal} \\
\mathbb{Z} & \mbox{otherwise},
\end{cases}
\end{equation}
while $\mathcal{Q}_p = 0$ for $p < -2$. We will not discuss the physical interpretations of these intrinsincally quasicrystalline topological phases in the present work very much, but it seems important to study them further.

\subsubsection{The general case (including non-invertible topological phases)}

Finally, we will describe the complete theory taking into account point-group symmetries, and applied also to non-invertible topological phases. Our main result is the \textbf{Quasicrystalline Equivalence Principle}, which states that the classification of quasicrystalline topological phases is in one-to-one correspondence with the classification of topological phases with only \emph{internal} symmetry, with an effective symmetry group $\widehat{G}$ related to the quasicrystalline structure. This is the quasicrystalline generalization of the ``Crystalline Equivalence Principle'' for crystalline topological phases proposed in Ref.~\cite{Thorngren_1612}.


\section{Crystalline SPTs and ``topological elasticity theory''}
\label{sec:crystal_elasticity}

In this section, we will review the classification of crystalline SPT phases for the case where there are no lattice symmetries other than translation symmetry, and the only internal symmetry is a $\UU(1)$ charge conservation symmetry. In this case, there is an integer valued invariant $\nu$, which represents the average charge per unit cell. The LSMOH theorem ensures that for a gapped SPT ground state, $\nu$ is an integer.

There is an important interpretation of this invariant $\nu$: it is the coefficient of a \emph{quantized topological term} that appears in the action describing the elasticity theory of the system
(i.e.\ the dynamics of phonons), after intergrating out all the degrees of freedom other than the long-wavelength elastic modes.
The elastic modes can be described by a slowly-varying
displacement field $\mathbf{u}(\mathbf{x},t)$ corresponding to the displacement of atoms from their equilibrium positions. The topological term is most naturally expressed in terms of the ``phase angles'' $\theta^I$ ($I = 1,\cdots,d$), defined in terms of $\mathbf{u}$ as
\begin{equation}
\label{eq:phase_defn_crystal}
    \mathbf{u}(\mathbf{x},t) = \frac{1}{2\pi} \mathbf{a}_I \theta^I(\mathbf{x},t) - \mathbf{x},
\end{equation}
where we use the repeated index summation convention, and $\mathbf{a}^I$ ($I=1,\cdots,d$) are a set of primitive lattice vectors. Then, in the presence of a background gauge field $A_\mu$ for the $\UU(1)$ symmetry, the Lagrangian can contain, in addition to the kinetic term, a topological term
\begin{align}
    \label{eq:c-topterm1}
    \mathcal{L} &= \mathcal{L}_0 + \frac{\nu}{2\pi} \epsilon^{\mu \nu} A_\mu \partial_\nu \theta^1 \quad (d = 1), \\
    \label{eq:c-topterm2}
    \mathcal{L} &= \mathcal{L}_0 + \frac{\nu}{8\pi^2} \epsilon^{\mu \nu \lambda} \epsilon_{IJ} A_\mu \partial_\nu \theta^I \partial_\lambda \theta^J, \quad (d=2) \\
    \label{eq:c-topterm3}
    \mathcal{L} &= \mathcal{L}_0 + \frac{\nu}{48\pi^3} \epsilon^{\mu \nu \lambda \sigma} \epsilon_{IJK} A_\mu \partial_\nu \theta^I \partial_\lambda \theta^J \partial_\sigma \theta^K \quad(d=3).
\end{align}
By demanding that the action be invariant (modulo $2\pi$) under large gauge transformations of $A$ on any space-time manifold, one can show that $\nu$ must be an integer.

From the topological term, one can derive various properties. Let us focus on the $d=2$ case for concreteness. Then, one finds that the charge density is
\begin{equation}
\label{eq:crystal_charge_density}
    \rho = \frac{\delta S}{\delta A_0} = \frac{\nu}{8\pi^2} \epsilon^{ij} \epsilon_{IJ} \partial_i \theta^I \partial_j \theta^J.
\end{equation}
In particular, in equilibrium one has (setting $\mathbf{u} = 0$ in \eqnref{eq:phase_defn_crystal}) that $\theta^I = K\indices{^I_i} x^i$ (where $K\indices{^I_i}$ is the inverse of the matrix $(1/2\pi) a\indices{^i_I}$ whose columns are the vectors $(1/2\pi) \mathbf{a}^I$), which gives
\begin{equation}
\label{eq:eq_rho_crystal}
    \rho = \frac{\nu}{4\pi^2} \det K = \frac{\nu}{V_{\mathrm{unit}}},
\end{equation}
where $V_{\mathrm{unit}} = \det a$ is the volume of the unit cell. Note that because the topological term is part of the effective theory that describes the system on length scales large compared with the unit cell size, the $\rho$ that we have calculated represents the \emph{average} charge density over such a scale; of course, the \emph{microscopic} charge density can vary rapidly over the scale of the unit cell, which is not captured by the effective theory.
From \eqnref{eq:eq_rho_crystal} one finds that the charge per unit cell is $\nu$, as expected. In fact, in general one can show that \eqnref{eq:crystal_charge_density} is equivalent to
\begin{equation}
    \rho = \frac{\nu}{\widetilde{V}_{\mathrm{unit}}},
\end{equation}
where $\widetilde{V}_{\mathrm{unit}}$ is the volume of the deformed unit cell due to the local strain.
One can also compute the current density
\begin{equation}
    J^i = \frac{\delta S}{\delta A_i} = \frac{\nu}{4\pi^2} \epsilon_{IJ} \epsilon^{ij} \partial_j \theta^I \partial_0 \theta^J,
\end{equation}
which one can convince oneself is indeed the correct expression for the current density for elastic deformations of a crystal in which a charge $\nu$ is bound to each unit cell.

Here, we should mention some conceptual slight-of-hand that has taken place. Normally one thinks of crystalline SPT phases characterized by the integer invariants $\nu$ as protected by the lattice translation symmetry. By contrast, phonons are the Goldstone modes arising from the spontaneous breaking of a continuous translation symmetry down to the aforementioned lattice translation symmetry. One might wonder what the topological term of the Goldstone modes has to do with the SPT phase with respect to the residual symmetry. However, in fact one does expect there to be a general relation \cite{Else_2102}.

The topological elasticity point of view proves particularly fruitful when one passes from crystals, which have discrete translation symmetry, to quasicrystals, which have \emph{no translation symmetry at all}. Nevertheless, what we will show is that the elasticity theory of quasicrystals still admits quantized topological terms, in fact with an even richer structure compared to the crystalline case, and that these topological terms correspond to distinct quasicrystalline phases of matter.

Finally, let us note that, if we want to take the view that we are classifying gapped topological phases of matter as described at the end of Section~\ref{subsec:defn_2}, we can view the elastic deformations described by the field $\theta^I$ as being imposed externally, through spatial and temporal variations in the pinning potential that we imposed to make the system gapped, in which case $\theta^I$ is a non-dynamical probe field. If, on the other hand, we do not care about precise definitions of topological phases and simply want to study the physical properties of a real crystal, we are free to treat the $\theta^I$ as gapless dynamical fields and study the physical implications of the topological terms. In the latter case, the prerequisite property for the validity of our discussion will be that the low-energy effective theory of the system should contain only the elastic modes; thus, metals for instance are excluded.


\section{Review of quasicrystals}
\label{sec:qcreview}

\subsection{Definition of quasicrystals and quasicrystalline topological phases}
\label{subsec:defn_2}

Let us describe the definition of a quasicrystal that we adopt in the present paper. We consider systems for which there is a countable set $\mathcal{L}$ of wavevectors such that the expectation value $\langle \hat{o}(\mathbf{x}) \rangle$ of all local observables in the ground state can be expanded in a Fourier series:
\begin{equation}
\label{eq:fourier_observable}
\langle \hat{o}(\boldsymbol{x}) \rangle = \sum_{\mathbf{Q} \in \mathcal{L}} a_{\mathbf{Q}} \exp(i \mathbf{Q} \cdot \boldsymbol{x}),
\end{equation}
with coefficients $a_{\mathbf{Q}}$ that depend on the choice of operator $\hat{o}$.
In a crystal, $\mathcal{L}$ is simply the reciprocal lattice of the crystal, and for a system in $d$ spatial dimensions it can be generated by a finite set of $d$ vectors (the primitive reciprocal lattice vectors), in the sense that all elements of $\mathcal{L}$ can be written as an integer linear combination of primitive reciprocal lattice vectors. By contrast, we say that the system is a \emph{quasicrystal} if $\mathcal{L}$ can be generated by a finite set of vectors, but the smallest such set has size greater than $d$.

When we talk about a quasicrystalline topological phase, we will mean a family of Hamiltonians such that the ground state is always gapped and satisfies the quasicrystallinity property stated above, \emph{with the reciprocal lattice $\mathcal{L}$ held fixed throughout the whole family}. In the crystalline case, this would be equivalent to demanding that the ground state always respects the discrete translation symmetry of the lattice i.e., to a topological phase protected by discrete translation symmetry. We will return to the question of how to consider topological phases protected by the point-group ``symmetry'' of a quasicrystal later on. We also note that the definition of quasicrystalline topological phases is distinct from that of ``local isomorphism" classes of quasicrystals -- see Appendix~\ref{sec:LI} for a discussion.

\subsection{Phonons and phasons in quasicrystals}
\label{subsec:phason}

The main reason that the elasticity theory of quasicrystals is richer than that of crystals is that there are two independent types of elastic deformations in a quasicrystal, corresponding to ``phonon'' and ``phason'' modes. In this section, we will review the physics of phonons and phasons, with a particular view to setting up the notation that we will use in later sections.

Let us imagine a ground state in which the local observables satisfy \eqnref{eq:fourier_observable}. One can then argue that, if the Hamiltonian has continuous translation symmetry (that is, we are thinking of the quasicrystal as resulting from spontaneously breaking continuous translation symmetry), then there is a whole manifold of ground states with equal energy in the thermodynamic limit.
Specifically, one can show that there is a ground state in which expectation values take the form
\begin{equation}
\label{eq:fourier_observable_prime}
\langle \hat{o}(\mathbf{x}) \rangle' = \sum_{\mathbf{Q} \in \mathcal{L}} a_{\mathbf{Q}} \exp(i\phi_{\mathbf{Q}} + i \mathbf{Q} \cdot  \mathbf{x}),
\end{equation}
with the same coefficients $a_{\mathbf{Q}}$ as the original state,
for any choice of phases $\phi_{\mathbf{Q}}$ (\emph{independent of} $\hat{o}$) satisfying
\begin{equation}
\label{eq:phi_homomorphism}
\phi_{\mathbf{Q}_1 + \mathbf{Q}_2} = \phi_{\mathbf{Q}_1} + \phi_{\mathbf{Q}_2}, \quad [\mathrm{mod}\, 2\pi]
\end{equation}
for any $\mathbf{Q}_1, \mathbf{Q}_2 \in \mathcal{L}$. The proof is basically that a state that assigns expectation values \eqnref{eq:fourier_observable_prime} is either a translation of the original state, or else can be \emph{arbitrarily well locally approximated} by translations of that state (in the language of Appendix \ref{sec:LI}, it is in the same ``local isomorphism class'' as the original state); either way, it must have the same energy.

The phases satisfying \eqnref{eq:phi_homomorphism} can be thought of as the ``order parameter'' that labels the spontaneous symmetry breaking ground states. The low energy elastic deformations correspond to long-wavelength fluctuations of this order parameter.
Thus, we replace $\phi_{\mathbf{Q}}$ in \eqnref{eq:fourier_observable_prime} with slowly-varying functions $\phi_{\mathbf{Q}}(\mathbf{x},t)$ (which are still required to satisfy \eqnref{eq:phi_homomorphism} at each point $\mathbf{x}$). A helpful way to formulate this is by introducing the phase fields $\theta_{\mathbf{Q}}(\mathbf{x},t) = \phi_{\mathbf{Q}}(\mathbf{x},t) + \mathbf{Q} \cdot \mathbf{x}$; then we have
\begin{equation}
\label{eq:rho_in_terms_of_theta}
\langle \hat{o}(\mathbf{x}) \rangle'  = \sum_{\mathbf{Q} \in \mathcal{L}} a_{\mathbf{Q}} \exp[i \theta_{\mathbf{Q}}(\mathbf{x},t)]
\end{equation}
where $\theta_{\mathbf{Q}}(\mathbf{x},t)$ satisfies \begin{equation}
\label{eq:theta_homomorphism}
\theta_{\mathbf{Q}_1 + \mathbf{Q}_2}(\mathbf{x},t) = \theta_{\mathbf{Q}_1}(\mathbf{x},t) + \theta_{\mathbf{Q}_2}(\mathbf{x},t), \quad [\mathrm{mod}\, 2\pi]
\end{equation}
and where, for a low-energy deformation, we must have
\begin{equation}
\label{eq:gradtheta}
    \nabla \theta_{\mathbf{Q}} (\mathbf{x},t) \approx \mathbf{Q},
\end{equation}
while in the ground state \eqnref{eq:gradtheta} becomes an equality.

We can obtain a concrete parameterization of the solutions to \eqnref{eq:theta_homomorphism} by introducing a set $\mathbf{K}^{1}, \cdots, \mathbf{K}^D$, of reciprocal vectors that generate $\mathcal{L}$. For a crystal one would have $D = d$, whereas for a quasicrystal $D > d$.
 Specifically, we impose the following properties:
\begin{enumerate}
\item \label{it:first}Every $\mathbf{Q} \in \mathcal{L}$ can be written as an integer linear combination of the $\mathbf{K}^I$'s.
\item \label{it:second} The $\mathbf{K}$'s are linearly independent over the integers; if $ n_I \textbf{K}^I = 0$ for some integers $n_I$ (here we are using the repeated index summation convention), then all the $n_I$'s are zero.
\end{enumerate}
In some cases, it is traditional to drop the second requirement and have an overcomplete set of vectors. This is commonly done in order for the set of generating vectors to be closed under point-group symmetry, as with the pentagonal quasicrystal discussed in Ref.~\cite{levine1985}, for which an overcomplete set of five vectors is traditionally used even though one of them is an integer linear combination of the other four. However, in this paper we will not adopt such an approach and our generating sets will never be overcomplete.
Note that because there are infinitely many possible choices of generating sets for a given $\mathcal{L}$, related by $\mathrm{GL}(D,\mathbb{Z})$ transformations, there will always be a $\mathrm{GL}(D,\mathbb{Z})$ gauge freedom associated with formulas that depend on the definition of the $\mathbf{K}^I$'s. In Appendix \ref{sec:octagonal}, we discuss an example of the choice of $\mathbf{K}$ vectors for an octagonal quasicrystal in 2D.

The solutions to \eqnref{eq:theta_homomorphism} can then be parameterized in terms of $D$ phase angle fields $\theta^I(\mathbf{x})$ ($I = 1,\cdots,D$) as:
\begin{equation}
\theta_{\mathbf{Q}}(\mathbf{x},t) = \theta^I(\mathbf{x},t) n_I(\mathbf{Q})
\end{equation}
where $n_I(\mathbf{Q})$ is the unique integer vector such that $\mathbf{Q} = n_I \mathbf{K}^I$.  \eqnref{eq:gradtheta} can then be expressed as
\begin{equation}
    \nabla \theta^I(\mathbf{x},t) \approx \mathbf{K}^I.
\end{equation}
Thus, the elastic deformations are parameterized by the $D$ fields $\theta^I(\mathbf{x},t)$. In a crystal, we have that $D = d$, and these are just the usual phonon modes. On the other hand, for a quasicrystal, we have $D > d$ and there are more elastic modes than in a crystal. It is conventional~\cite{lubensky1985,levine1985,lubensky1986,lubensky1986b,ding1993} to decompose these modes into ``phonon'' and ``phason'' modes, corresponding to certain linear combinations of the $\theta^I$'s. However, in this paper we will find it more convenient to work with the $\theta^I$'s directly.

Finally, let us note an appealing interpretation of the $\theta$ fields. Define a function $\widetilde{o}$ on $\mathbb{R}^D$ according to
\begin{equation}
    \widetilde{o}(\boldsymbol{\theta}) = \sum_{\mathbf{n} \in \mathbb{Z}^D} a_{n_I \mathbf{K}^I} \exp(i \mathbf{n} \cdot \boldsymbol{\theta}),
\end{equation}
which is periodic with respect to $2\pi$ translations along any of the coordinate directions in $\mathbb{R}^D$, and where $a_{\mathbf{Q}}$ are the Fourier coefficients in \eqnref{eq:fourier_observable}. Then we can write \eqnref{eq:fourier_observable_prime} as
\begin{equation}
   \langle\hat{o}(\mathbf{x})\rangle' = \widetilde{o}(\boldsymbol{\theta}(\mathbf{x},t)),
\end{equation}
where $\boldsymbol{\theta}(\mathbf{x},t)$ is the vector in $\mathbb{R}^D$ whose components are $\theta^I(\mathbf{x},t)$. Thus, in the ground state, where $\mathbf{\theta}^I(\mathbf{x},t) = \theta^I_{(0)} + \mathbf{K}^I \cdot \mathbf{x}$, the expectation values of observables in a quasicrystal is given by linearly mapping $d$-dimensional physical space into a hyperplane slicing through a $D$-dimensional ``crystal. Moreover, phonon  and phason deformations correspond to keeping the $D$-dimensional ``crystal'' fixed (so the function $\widetilde{o}$ remains fixed), but deforming the mapping from $d$-dimensional space to the $D$-dimensional ``superspace.''


\section{Topological term for the elasticity theory of quasicrystals}
\label{sec:qc_topoterm}

The generalization to quasicrystals of the topological term discussed in Section \ref{sec:crystal_elasticity}, expressed in terms of the fields $\theta^I$ introduced in Section~\ref{subsec:phason}, can be written as:
\begin{align}
\label{eq:first_topterm}
\mathcal{L} &= \mathcal{L}_0 + \frac{1}{2\pi}  C_I \epsilon^{\mu \nu} A_\mu \partial_\nu \theta^I, \quad (d=1) \\
\mathcal{L} &= \mathcal{L}_0 + \frac{1}{8\pi^2} C_{IJ} \epsilon^{\mu \nu \lambda} A_\mu \partial_\nu \theta^I \partial_\lambda \theta^J, \quad (d=2) 
\label{eq:2ndtopterm}
\\
\mathcal{L} &= \mathcal{L}_0 + \frac{1}{48\pi^3} C_{IJK} \epsilon^{\mu \nu \lambda \sigma} A_\mu \partial_\nu \theta^I \partial_\lambda \theta^J \partial_\sigma \theta^K. \quad (d=3)
\label{eq:lasttopterm}
\end{align}
Here, $C_I$ is a $D$-dimensional vector, $C_{IJ}$ is a $(D \times D)$  antisymmetric matrix, and $C_{IJK}$ is a rank-3 tensor of dimension $D$ which is antisymmetric in all pairs of indices.
By demanding that the action be invariant (modulo $2\pi$) under large gauge transformations of $A$ on any space-time manifold, one can show that each entry of $C$ is quantized to be an integer.
Therefore, we obtain a partial classification of quasicrystalline SPTs with $\mathrm{U}(1)$ symmetry by integral antisymmetric rank-$d$ tensors of dimension $D$. In the crystalline case, where $D=d$, all such tensors are simply integer multiples of the Levi-Civita tensor, and we recover \eqnref{eq:cryst_topterm} in $d=2$ for example.

Note that, for a given quasicrystalline SPT phase with underyling reciprocal lattice $\mathcal{L}$, the entries of $C$ depend on the arbitrary choice of the generating reciprocal vectors $\mathbf{K}^I$'s for $\mathcal{L}$. We will see later that a more abstract way to state the classification, that has the advantage of being gauge-invariant, is as follows: the SPT phases are classified by integral cohomology $H^d(\mathcal{L}^*, \mathbb{Z})$, where $\mathcal{L}$ is the space of homomorphisms from $\mathcal{L}^*$ into $\UU(1)$; that is, the space of solutions to \eqnref{eq:phi_homomorphism} (note that we are taking singular cohomology of the topological space $\mathcal{L}^*$, discarding its group structure, \emph{not} group cohomology). We can recover the concrete classification in terms of integral antisymmetric tensors if we observe that $\mathcal{L}^*$ is topologically a $D$-torus.

For quasicrystals, there are in fact additional possible topological terms that one can write, which do not depend on the $\mathrm{U}(1)$ symmetry. Firstly, if $D \geq d+1$, then we can write the term
\begin{align}
\label{eq:first_theta}
    \mathcal{L}_{\mathrm{Theta}} &= \Theta_{IJ} \frac{1}{8\pi^2} \epsilon^{\mu \nu} \partial_\mu \theta^I \partial_\nu \theta^J \quad (d=1), \\
    \mathcal{L}_{\mathrm{Theta}} &= \Theta_{IJK} \frac{1}{48\pi^3} \epsilon^{\mu \nu \lambda} \partial_\mu \theta^I \partial_\nu \theta^J \partial_\lambda \theta^J \quad (d=2), \label{eq:second_theta}
\end{align}
and so on, where $\Theta$ is an antisymmetric tensor in dimension $D$. One can show that if the entries of $\Theta$ are integer multiples of $2\pi$, then the action is a multiple of $2\pi$ on any closed manifold; therefore, the entries of $\Theta$ are only defined modulo $2\pi$. Hence, since under time-reversal symmetry we have $\Theta \to -\Theta$, we see that if we impose time-reversal symmetry then the components of $\Theta$ are quantized to be $0$ or $\pi$ (modulo $2\pi$). Thus, for each independent component of $\Theta$, we obtain a $\mathbb{Z}_2$ quantized topological invariant (these arguments should, of course be very reminiscent of those for a topological insulator in three spatial dimensions~\cite{Qi_0802}). In the more abstract language, these topological terms (and hence, the corresponding quasicrystalline SPTs) are classified by $H^{d+1}(\mathcal{L}^*, \mathbb{Z}_2)$.

Finally, if time-reversal symmetry is broken and $D \geq d+2$, then there is an additional class of topological terms of the Wess-Zumino type~\cite{AltlandSimons}. These ones are not canonically expressible as a local Langrangian in $d+1$ space-time dimensions. Instead, the action is written in terms of an extension of the $\theta$ fields into one higher dimension:
\begin{equation}
    S_{\mathrm{WZ}}[\theta] = \int_{M_{d+2}} \mathcal{L}_{d+2}^{\Theta} [\widetilde{\theta}],
\end{equation}
where $M_{d+2}$ is a $(d+2)$-dimensional manifold whose $(d+1)$-dimensional boundary corresponds to the physical space-time, and $\widetilde{\theta}$ is an extension of the $\theta$ fields onto $M_{d+2}$. $\mathcal{L}_{d+2}$ is an action of the form analogous to Eqs.~(\ref{eq:first_theta}) and (\ref{eq:second_theta}) (but in space-time dimension $d+2$), parameterized by a rank-$d+2$, dimension $D$ antisymmetric integer tensor. In order for the
action to be independent of the choice of extension $\widetilde{\theta}$, we require that all the components of $\Theta$ are an integer multiple of $2\pi$. In the abstract language, these terms are classified by $H^{d+2}(\mathcal{L}^*, \mathbb{Z})$.

In this paper, we will focus on the topological terms of the form Eqs.~(\ref{eq:first_topterm}--\ref{eq:lasttopterm}), leaving a detailed examination of the consequences of the other topological terms we have described for future work. 

\section{Mobility of dislocations in a quasicrystal}
\label{sec:mobility}

A well-known fact about defects in crystals is that dislocations can only move in the direction of their Burgers vector. Specifically, this applies to systems with a conserved charge, such that the charge $\nu$ per unit cell (see Section~\ref{subsec:elasticity}) is nonzero. One can then show that moving the dislocation in a direction not parallel to the Burgers vector would violate charge conservation. The approach based on topological terms in the elasticity theory has been given in Ref.~\cite{Cvetkovic_0508}.

In this section, we will show that a generalization of this mobility constraint applies to dislocations in two-dimensional quasicrystals. A dislocation in a quasicrystal is characterized by a non-trivial winding number of the $\theta$ field defined in Section~\ref{subsec:phason} above as the dislocation is encircled:
\begin{equation}
\frac{1}{2\pi} \oint_{C} dx^{i}\partial_{i}\theta^{I} = b^{I} \in \mathbb{Z}^{D},
\label{eqn:dislocation}
\end{equation}
where $C$ is a loop surrounding the dislocation, and $b^{I}$ is the Burgers vector. 
The topological mobility constraint of a dislocation can be derived from the charge conservation, as we now show.

From \eqnref{eq:2ndtopterm}, we see that there is a contribution to the current given by
\begin{equation}
J^{\mu} = \frac{1}{8\pi^2} C_{IJ} \epsilon^{\mu \nu \lambda} \partial_{\nu}\theta^{I} \partial_{\lambda}\theta^{J}.
\label{eqn:current}
\end{equation} 
From this, we find that the condition for local charge conservation takes the form
\begin{eqnarray}
\partial_{\mu}J^{\mu} = C_{IJ} \epsilon^{\mu \nu \lambda} (\partial_{\mu} \partial_{\nu} \theta^{I} - \partial_{\nu} \partial_{\mu} \theta^{I}) \partial_{\lambda} \theta^{J} = 0.
\label{eq:qccurrentcons}
\end{eqnarray}
Assuming that we are close to the equilibrium configuration, we have
\begin{eqnarray}
\partial_{0}\theta^{I} \approx 0, \quad
\partial_{i}\theta^{I} \approx K\indices{^{I}_{i}}.
\end{eqnarray}
We then expand Eq,~\eqref{eq:qccurrentcons} to the leading order in $\partial_{i}\theta^{I} - K\indices{^I_i}$, giving
\begin{equation}
\partial_{\mu}J^{\mu} = C_{IJ} \epsilon^{ij} (\partial_{0} \partial_{i} \theta^{I} - \partial_{i} \partial_{0} \theta^{I}) K\indices{^J_j} = 0
\end{equation}
Now consider a dislocation with Burgers vector $b^{I}$ at position $x^{i}(t)$ moving at velocity $v^{j}$.
Then one finds that
\begin{equation}
(\partial_0 \partial_i - \partial_i \partial_0) \theta^I = \epsilon_{ij} v^j \delta^2(\mathbf{x} - \mathbf{v}(t)),
\end{equation}
and hence
 the local charge conservation requires that
\begin{equation}
C_{IJ}b^{I}K\indices{^J_j} v^{j} = 0.
\label{eq:mobilityconst}
\end{equation}
Eq.~\eqref{eq:mobilityconst} gives the topological mobility constraint of dislocations in a two-dimensional quasicrystal. We see that a dislocation can only move along the direction set by a combination of the reciprocal lattice vectors, the SPT invariant $C_{IJ}$, and the Burgers vector $b^{I}$. 

One can, of course, raise an objection regarding the rigor of the above argument, since it is based on linearizing about the equilibrium configuration, while in fact $\theta^I$ will always become singular at the dislocation core. Moreover, one does does not expect the continuum field theory description that we are using to be valid near the dislocation core. In Appendix~\ref{appendix:mobility}, we give a more careful argument for the constraint~\eqnref{eq:mobilityconst} by invoking the continuum field theory description only far away from the dislocation core, where one expects such a description to be valid.

Note that there is always at least one direction $v^j$ that satisfies \eqnref{eq:mobilityconst}. Thus, as in crystals, dislocations in quasicrystals are never completely immobilized but can move in some direction. On the other hand, at low temperatures, quasicrystals are found experimentally to be quite brittle, reflecting the fact that for dynamical reasons, it is more difficult (but not impossible) for dislocations to move in quasicrystals than in crystals~\cite{socolar86}, even though there is no topological constraint that fully immobilizes them.

\section{Average charge density for quasicrystals and the LSMOH theorem}
\label{sec:lsmoh}

Another consequence of the topological terms Eqs.~(\ref{eq:first_topterm}--\ref{eq:lasttopterm}) is for the overall charge density of the system. Since the charge density can be evaluated as $\rho = \delta S/\delta A_0$, and using the fact that, in the ground state $\partial_i \theta^I = K\indices{^I_i}$, we find the average density
\begin{align}
\label{eq:rho_first}
    \rho &= \frac{1}{2\pi} C_I K\indices{^I}, \quad (d=1),\\
\label{eq:rho_second}
    \rho &= \frac{1}{8\pi^2} C_{IJ} \epsilon^{ij} K\indices{^I_i} K\indices{^J_j}, \quad (d=2), \\
    \rho &= \frac{1}{48\pi^2}  C_{IJK} \epsilon^{ijk} K\indices{^I_i} K\indices{^J_j} K\indices{^K_k} \quad (d=3) \, .
    \label{eq:rho_last}
\end{align}
Here, we have assumed that the non-topological part of the action does not contribute to the average charge density. This is a reasonable assumption, because the phonon and phason fields do not transform non-trivially under $\mathrm{U}(1)$, so they do not minimally couple to the gauge field (and one can verify that a non-minimal coupling would not give any contribution to the ground state charge density). 

In the crystalline case, where $C$ is always an integer multiple $\nu$ of the Levi-Civita symbol, then this simply amounts to saying that the charge per unit cell is $\nu$. More generally, a helpful way to interpret these expressions is in terms of the tile picture of Section~\ref{sec:tiling}; these densities are precisely the average charge densities that one would expect from having integer charges bound to the tiles (with the charge bound to each class of tiles given by the appropriate entry of $C$), once one takes into account how often each tile class appears in a tiling of space.

The set $\mathcal{S}(\mathcal{L})$ of all allowed densities, generated by varying $C$ in Eqs.~(\ref{eq:rho_first}--\ref{eq:rho_last}) over all integer anti-symmetric tensors, determines the quasicrystalline version of the LSMOH theorem: only densities in this set are allowed in an insulating quasicrystal with reciprocal lattice $\mathcal{L}$ without non-invertible topological order (i.e., without ground state degeneracy on the torus).
Unlike in the crystalline case, the set $\mathcal{S}(\mathcal{L})$ is typically \emph{dense} in the space of all real numbers; that is, for any $\rho \in \mathbb{R}$ we can find allowed densities arbitarily close to $\rho$.
However, the allowed densities are not \emph{continuous} -- there is no interval $
[\rho_0, \rho_1]$ with $\rho_0 < \rho_1$ such that $[\rho_0, \rho_1]$ is a subset of $\mathcal{S}(\mathcal{L})$ -- as can be seen from the fact that $\mathcal{S}(\mathcal{L})$ is a countable set. Note that in order for $\mathcal{S}(\mathcal{L})$ to be dense, one must allow the integer entries of $C$ to take arbitrary large values. In fact, we expect that systems with large entries of $C$ are probably very hard to realize in practice.

Finally, let us remark that in the case of \emph{free-fermion} insulators, the formulas Eqs.~(\ref{eq:rho_first}--\ref{eq:rho_last}), involving the quantized integer ``gap labels'' $C$, have long been known \cite{Bellissard1,Bellissard2}. Our work can thus be seen as a demonstration that these free-fermion topological invariants are robust to arbitrary local interactions that respect the quasicrystalline structure.

\section{Simple Interpretations of the SPT Invariants}
\label{sec:interpret}

\subsection{Tiling Interpretation}
\label{sec:tiling}

We now discuss a simple interpretation of the SPT invariants $C$ in Eqs.~\eqref{eq:first_topterm}-\eqref{eq:lasttopterm} in terms of the tiles forming the quasicrystal. Recall first the well-known fact that a quasicrystalline structure can be obtained by tiling space in such a way that each tile $T_i$ can, by translation, be related to one of a finite set $\{T_i \}_{i=1}^n$ of primitive tiles. In fact, there is a canonical way to generate such a tiling for any reciprocal lattice $\mathcal{L}$ \cite{Oguey1988}, such that for an $\mathcal{L}$ with $D$ primitive vectors, there are $n = {D \choose d}$ primitive tiles. In the case of a periodic crystal, we have $D=d$ and hence $n = 1$, so that only a single primitive tile is required (the unit cell).

\begin{figure}
\includegraphics[width=0.8\columnwidth]{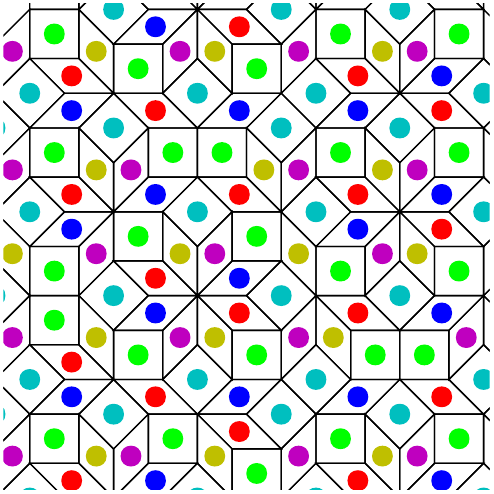}
\caption{\label{fig:weakdots} A construction of the $k=2$ quasicrystalline topological phases for a $d=2$, $D=4$ quasicrystal. Each color (of which there are 6) is assigned a topological invariant that can be chosen independently (if no rotational symmetry is imposed) and represents the charge bound to the corresponding tile. If the 8-fold rotational symmetry is imposed, then there are two independent invariants, corresponding to the square and rhombus tiles.}
\end{figure}

Now, since $C$ is a rank-$d$ anti-symmetric integer tensor of dimension $D$, it has ${D \choose d}$ independent entries. The basic idea is that, in a suitable limit, we can interpret these integer entries of $C$ as the charge bound to the corresponding tiles of the quasicrystal. For example, Figure \ref{fig:weakdots} illustrates this picture for the case of a $d=2$, $D=4$ quasicrystal, for which there are $6$ independent invariants (for a discussion of how the tiling shown in Figure \ref{fig:weakdots}, known as the Ammann-Beenker tiling, is constructed, see Appendix \ref{sec:abtiling}).
In the case $D=d$, where there is only one kind of tile, this evidently reduces to just the total charge per unit cell. Of course, this picture is only valid in the limit where the charges are tightly bound to the tiles; however, the discussion in the previous section demonstrates that the SPT invariants are still robust in the presence of charge fluctuations, as long as the system remains gapped.

\begin{figure}
\includegraphics[width=0.8\columnwidth]{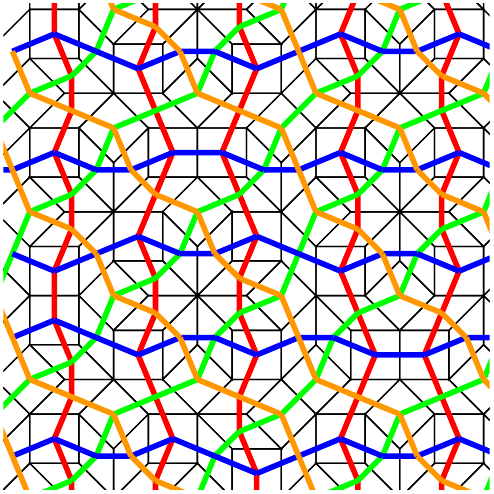}
\caption{\label{fig:weak1d} A construction of the $k=1$ 
quasicrystalline topological phases for a $d=2$, $D=4$ quasicrystal. Each color (blue, orange, green, red) is assigned a 1D SPT invariant that can be chosen independently (if no rotational symmetry is imposed) and represents the 1D SPT phases placed along the corresponding paths. If the 8-fold rotational symmetry is imposed, there is only one independent SPT invariant.}
\end{figure}

This picture corresponds to the $k=d$ invariants in the general classification discussed in Section \ref{subsubsec:weak}. The picture can also be generalized to the other SPT invariants in that classification, at least for $k \leq d$. For example, consider the case of $k=d-1$. Then there are ${D \choose d-1}$ independent SPT invariants, each of which takes values in the classification group of $1$-dimensional SPTs with respect to the internal symmetry. ${D \choose d-1}$ is the number of different kinds of codimension-1 faces of the tiles. Thus the SPT states correspond to piercing the codimension-1 faces with the corresponding 1-dimensional SPTs. This is illustrated in Figure \ref{fig:weak1d}.

In order to demonstrate the validity of this tile picture in the $k=d$ case for a system with $\mathrm{U}(1)$ symmetry, it is instructive to compute the average charge density $\rho$ of the system. Let us focus on the $d=2$ case for concreteness. The tiles' classes are then labelled by pairs of indices $(IJ)$. Then, since each tile of type $(IJ)$ carries charge $C_{IJ} \Sigma_{IJ}$ (where $\Sigma_{IJ} = \pm 1 $ defines some sign convention, such that $\Sigma_{IJ} = -\Sigma_{JI}$), we find that
\begin{equation}
\rho = \frac{1}{2} C_{IJ} \rho^{IJ},
\end{equation}
where $\rho^{IJ} = \Sigma_{IJ} |\rho^{IJ}|$, and $|\rho^{IJ}|$ is the average number of $(IJ)$-type tiles in the tiling per unit volume.
Moreover, we show in Appendix \ref{appendix:tile_density} that
\begin{equation}
\label{eq:rhoIJ2d}
\rho^{IJ} = \frac{1}{(2\pi)^2}K\indices{^I_i} K\indices{^J_j} \epsilon^{ij}.
\end{equation}
Hence, we find
\begin{equation}
\rho = \frac{1}{8\pi^2} C_{IJ} K\indices{^I_i} K\indices{^J_j} \epsilon^{ij}.
\end{equation}
which agrees with \eqnref{eq:rho_second}.

\subsubsection*{Example: Fibonacci tiling}

\begin{figure}[t]
    \centering
    \includegraphics[width=0.5\textwidth]{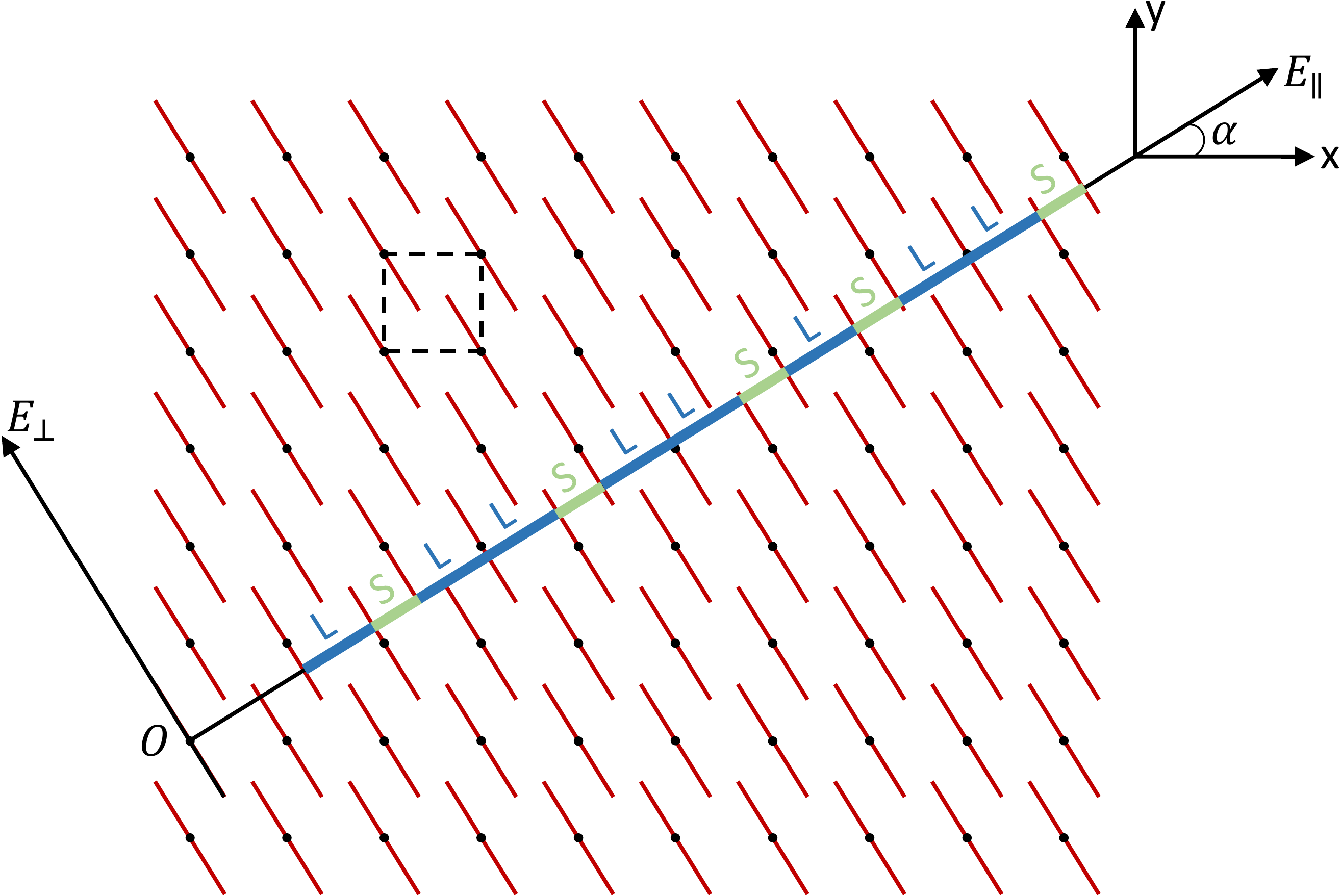}
    \caption{Generating the $d=1$ Fibonacci quasicrystal from a $D=2$ square lattice decorated with $d=1$ atomic surfaces $\sigma$ (red line segments). Black dots denote lattice sites and dashed black lines indicate the superspace unit cell. Intersections between the atomic surfaces and physical space $\Epar$ generate the quasicrystal; the intersections divide $\Epar$ into a sequence of long (L, in blue) and short (S, in green) intervals. 
    \label{fig:tile}}
\end{figure}

To further illuminate the tile picture (for the $k=d$ case), let us consider for simplicity the case of a Fibonacci quasicrystal in $d=1$ dimension. In this case, we have $D=2$ and so ${D \choose d} = 2$ distinct tiles. A typical Fibonacci quasicrystal is obtained by putting particles at positions
\begin{equation}
x_{n} = x_{0} + n(3/\tau -1) +  (1/\tau - 1) \text{Frac}(n/\tau),
\end{equation}
where $\text{Frac}(x)$ is the fractional part of $x$ and $\tau = (1+\sqrt{5})/2$ is the golden ratio. The distances of neighboring points $x_{n}-x_{n-1}$ are either $1$ or $1/\tau$. One can thus equivalently view the Fibonacci quasicrystal as consisting of a sequence of intervals with length $1$ and $1/\tau$. These two types of intervals will be called a long (L) and a short (S) tile respectively---a typical Fibonacci quasicrystal is shown in Fig.~\ref{fig:tile}, where the L and S tiles are colored in blue and green respectively. The primitive reciprocal vectors $K^1$, $K^2$ (actually just scalars since $d=1$) for the Fibonacci quasicrystal are given by~\cite{Janssen2007}
\begin{equation}
\frac{1}{2\pi}\begin{pmatrix} K^1 \\ K^2\end{pmatrix} = \frac{1}{2-1/\tau} \left(
\begin{array}{c}
1 
\\
1/\tau 
\end{array}
\right).
\end{equation}
From Eq.~\eqref{eq:rho_first}, we obtain the charge density of the Fibonacci quasicrystal:
\begin{eqnarray}
\rho &=& \frac{1}{2 \pi} C_{I}K^{I}
\\
&=& \frac{C_{1}}{2-1/\tau} + \frac{C_{2}/\tau}{2-1/\tau}.
\label{eqn:rho-fibo}
\end{eqnarray}

We now show that we can interpret the SPT invariants $C_{1}$ and $C_{2}$ as the integer charges bound to the L and S tiles respectively. Consider a large region with length $V$ in the physical space, and let $C_{1}$ and $C_{2}$ be the integer charges bound to the L and S tiles respectively. The total charge $Q$ inside the region $V$ is given by 
\begin{equation}
Q = n_{L} C_{1} + n_{S} C_{2} ,
\end{equation}
where $n_{L}$ and $n_{S}$ denote the number of L and S tiles inside the region $V$. Let $l_{L}$ and $l_{S}$ be the lengths of L and S tiles. The total length of this region is then given by 
\begin{equation}
V = l_{L}n_{L} +l_{S}n_{S}.
\end{equation}
Combining the above, we find the average charge density:
\begin{eqnarray}
\frac{Q}{V} &=& \frac{n_{L} C_{1} + n_{S} C_{2}}{l_{L}n_{L} +l_{S}n_{S}}
\\
&=& \frac{C_{1} + C_{2}/\tau}{2-1/\tau} \frac{1}{l_{L}}.
\label{eqn:avg-charge}
\end{eqnarray}
To obtain the second equality, we have used the fact that the L tile appears more frequently than the S tile, and that the ratio of their frequencies converges to the golden ratio $\tau$. Hence, we have $n_{L} = \tau n_{S}$. We have also used that the lengths of the tiles satisfy $l_{L} = \tau l_{S}$ and $1+(1/\tau)^{2} = 2-1/\tau$. In calculating the average charge density, we see that it was important to take the frequency of each class of tiles into account, which is a unique feature of quasicrystals. Comparing Eq.~\eqref{eqn:avg-charge} with Eq.~\eqref{eqn:rho-fibo}, we further see that the density $\rho$ given in Eq.~\eqref{eqn:rho-fibo} is the average charge density over a scale much larger than the length of the L tile. We can therefore interpret the SPT invariants $C_{1}$ and $C_{2}$ as the integer charges bounds to the L and S tiles.

\subsection{Atomic surfaces}
\label{sec:intersection}

In this section, we provide another interpretation of the SPT invariants introduced in Eqs.~\eqref{eq:first_topterm}-\eqref{eq:lasttopterm}. Specifically, we will describe how to assign a quantized topological invariant to any quasicrystal such that the location of particles whose number is conserved forms a classical point set (they no longer need to rigidly bound to the center of tiles as they were in the previous interpretation). In this context, the invariant is a well-known one in the quasicrystal literature~\cite{Kalugin_1989, numbertheory92}.
In this section, we will refer to the particles as ``atoms'' in line with previous literature. As always, the considerations of the present paper demonstrate that the SPT invariants do remain well-defined beyond the limit of classical point particles, but in the current section we focus on this limit.

First, we need to introduce some important concepts of quasicrystals from the superspace perspective. Recall that a quasiperiodic arrangement of point atoms can be described as a $d$-dimensional section of a $D$-dimensional periodic measure (where there exists a unique minimal value for $D$~\cite{baake91}). More specifically, a function of $d$ real variables on an affine $d$-dimensional space $\Epar$ is said to be quasiperiodic if it is the restriction to $\Epar$ of some periodic function of $D$ real variables defined in a higher dimensional space $\mathbb{R}^D$. Of course, when the physical space $\Epar$ (also referred to as the ``embedded" space or the ``cut") is oriented rationally with respect to the lattice of periods of the periodic function, the restriction of this function to $\Epar$ is itself a periodic function. However, a quasiperiodic function is obtained upon restriction of the periodic function to a cut that is irrationally oriented.

As discussed in Ref.~\cite{numbertheory92}, to describe a quasiperiodic arrangement of point atoms, consider a $D$-dimensional periodic measure $\mu$ such that its restriction to $\Epar$ results in a quasiperiodic configuration of Dirac delta functions. Let $\Lambda$ be the lattice of periods of this measure; then, the support of $\mu$ defines a $\Lambda$-periodic geometric locus $\Sigma$. 
We refer to $\Sigma$ as the set of atomic surfaces.
The quasiperiodic atomic configuration is then generated by the intersection points of $\Sigma$ with $\Epar$.

As an example, let us consider a two-dimensional ($D=2$) square lattice (with lattice constant unity) and $\Epar$ given by a $d=1$ subspace oriented at an angle $\alpha$ with respect to the $x$-axis (see Fig.~\ref{fig:tile}). Each vertex of the square lattice is decorated with an atomic surface  of length $\ell = \sin(\alpha) + \cos(\alpha)$.
 The intersection points of $\Sigma$
 with $\Epar$ generate a quasicrystalline structure iff $\tan \alpha$ is irrational; in particular, one obtains the paradigmatic Fibonacci quasicrystal when $\tan(\alpha) = 1/\tau$, where $\tau = (1+\sqrt{5})/2$ is the golden ratio. As shown in Fig.~\ref{fig:tile}, $\Sigma$ divides $\Epar$ into a sequence of long (L) and short (S) intervals---or ``tiles"---of length $1$ and $1/\tau$ respectively.

\begin{figure} 
\includegraphics[width=0.8\columnwidth]{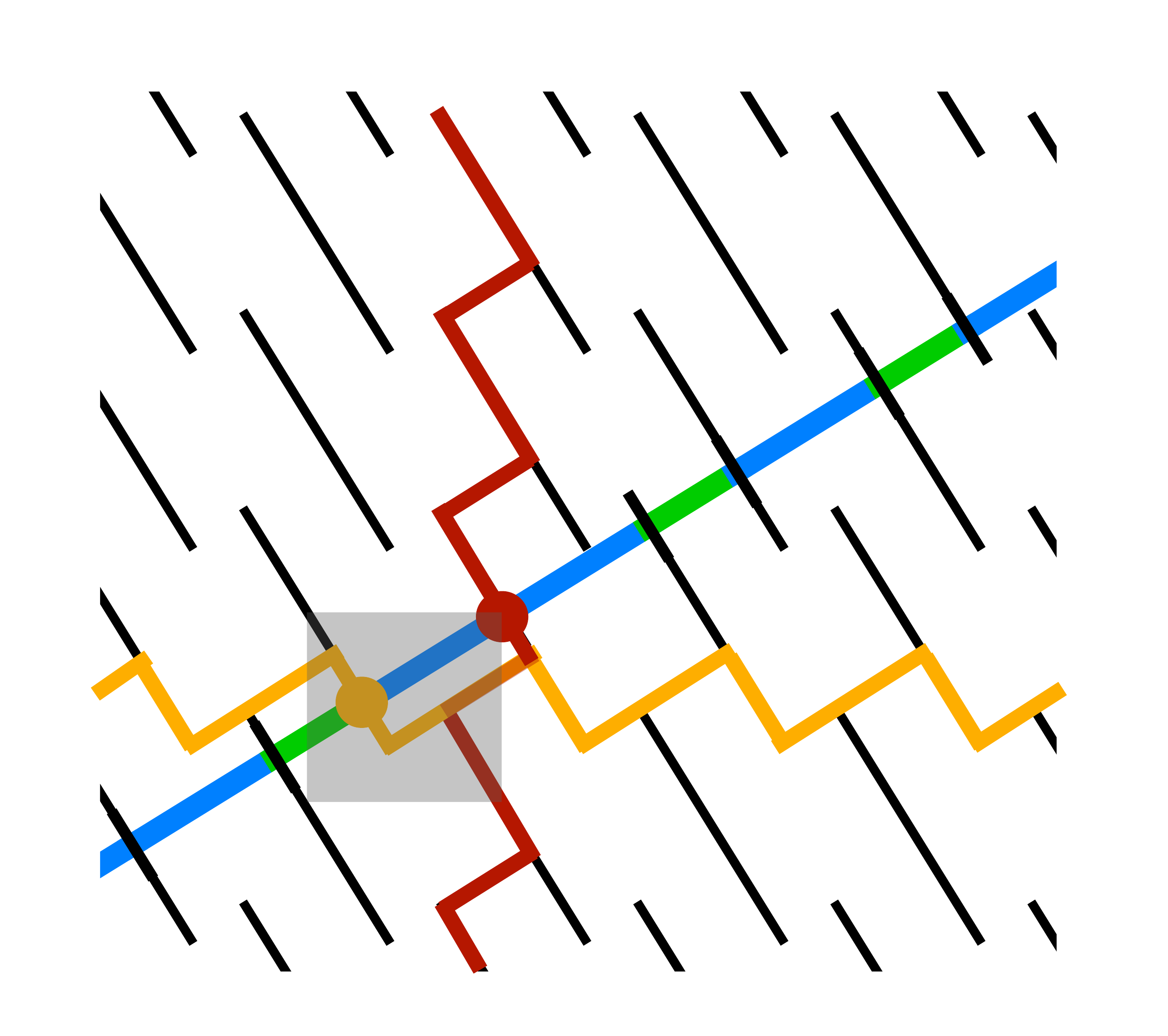}
\caption{Atomic surfaces (black lines) of a Fibonacci quasicrystal. The vertices of the Fibonacci quasicrystal is determined by the intersection points of the physical space and the atomic surfaces. The blue and green regions are the long (L) and the short (S) tiles that tile the $d=1$ physical space. The yellow line is an S-type atomic surface associated to the yellow particle. The red line is an L-type atomic surface associated to the red particle.
}
\label{fig:AS}
\end{figure}

We can imagine that there are two different species of atoms, corresponding to particles that appear to the right of an L tile  and atoms that appear to the right of an S tile; we refer to these as L and S atoms. Therefore, we can classify segments of atomic surface into two different ``types,'' depending on which kind of particle they produce upon intersection with $\mathbf{E}_\parallel$, as shown in Figure \ref{fig:AS}. Moreover, while the atomic surfaces are originally discontinuous, we will introduce segments that are parallel to the physical space to make the atomic surfaces of each type closed curves without affecting the quasicrystalline structure on $\mathbf{E}_\parallel$. 
The property that the atomic surfaces should be closed is referred to as the ``closeness condition''~\cite{numbertheory92}. Physically, this condition reflects the conservation of the two species of atoms, since it ensures that a phason deformation (which corresponds to moving $\mathbf{E}_\parallel$ in the superspace) at most causes particles to jump to nearby positions, but not to be created or destroyed. Henceforth, we will take the term ``atomic surface" to mean a continuous atomic surface.

Fig.~\ref{fig:AS} shows an S atom (yellow dot) with the corresponding S-type atomic surface represented by the yellow line. Fig.~\ref{fig:AS} also shows an L atom (red dot), with the corresponding S-type atomic surface shown in red.
As one can see from Fig.~\ref{fig:AS}, the S-type atomic surface winds non-trivially on the $2$-torus along the $\theta^{1}$-direction; that is, it has a ``winding number'' (1,0), while the S-type atomic surface winds non-trivially on the $2$-torus along the $\theta^2$ direction; that is, it has a winding number (0,1). These winding numbers precisely determine the SPT invariants with respect to the two $\mathrm{U}(1)$ symmetries corresponding to the conservation of S and L atoms respectively.

Similarly, one can give a general statement of the topological invariant for any configuration of atomic surfaces corresponding to a conserved species of particles. We represent an atomic surface as a $\mathbb{Z}$-valued $(D-d)$-cycle on $\mathbb{T}^{D}$. Topologically non-trivial configurations of the atomic surfaces are then classified by the homology group $H_{D-d}(\mathbb{T}^{D},\mathbb{Z})$. By applying the Poincaré duality, we see that these are then equivalently classified by the cohomology group $H^{d}(\mathbb{T}^{D},\mathbb{Z})$.
As discussed in Sec.~\ref{sec:qc_topoterm}, the SPT invariants are classified by the same cohomology group $H^{d}(\mathcal{L}^*,\mathbb{Z})$, with $\mathcal{L}^*$ topologically equivalent to $\mathbb{T}^D$;  therefore, the SPT invariants are determined by the (co)homology classes of the atomic surfaces on the $D$-torus.

This concludes our second interpretation of the SPT invariants $C$ in terms of the atomic surfaces. 
We finally note that this picture can be generalized in an obvious way to the other SPT invariants discussed in Section~\ref{subsubsec:weak} with $k < d$, where we replace the assumption of classical point particles in $d$ spatial dimensions with the assumption that there is some configuration of $(d-k)$-dimensional surfaces in $d$ spatial dimensions that host $(d-k)$-dimensional SPTs.


\section{The general classification of quasicrystalline topological phases}
\label{sec:general}

\subsection{Without point-group symmetry}
\label{sec:nopgs}

As we discussed in Section~\ref{subsec:phason} above, the ``order parameter manifold'' for a quasicrystal is $\mathcal{L}^*$, the group of homomorphisms from the reciprocal lattice $\mathcal{L}$ into $\mathrm{U}(1)$. As previously discussed, for example in Ref.~\cite{Else_2102}, in general ``topological terms for Goldstone modes'' arise when the symmetry-breaking ground states, viewed as a family of gapped ground states parameterized by the order parameter manifold, are a topologically non-trivial family. Thus, in order to classify quasicrystalline topological phases, which we view as being characterized by the topological term of the phonons and phasons, we want to classify topological families of ground states parameterized by $\mathcal{L}^*$.

Let us briefly note one subtlety, which is that for quasicrystals, the ground states are not necessarily continuous functions of the order parameter. For example, in Figure~\ref{fig:tile}, varying the order parameter would correspond to moving the location of the cut in superspace, and one readily sees that this will sometimes result in discontinuous jumps of the locations of particles. This is a well-known fact about phasons in quasicrystals. Nevertheless, these jumps always involve only local rearrangements; this implies, for example, that they can be effected by a local unitary. We conjecture, therefore, that the classification of topological families remains unchanged. In any case, these discontinuities are likely an artifact of treating the particle locations as totally classical variables, and would likely get smoothed out once one allows these variables to quantum fluctuate.

The problem of classifying topological classes of families of gapped ground states parameterized by a space $\mathcal{L}^*$, in the presence of internal symmetry $G_{\mathrm{int}}$, has previously been considered in Refs.~\cite{Thorngren_1612, Thorngren_1710, Gaiotto_1712, Cordova_1905_a, Cordova_1905_b, Kapustin_2001, Kapustin_2003, Hsin_2004}. Although not rigorously proven, it is believed that for \emph{invertible} states, such families are classified by $h^d(\mathcal{L}_* \times BG_{\mathrm{int}})$, where $BG_{\mathrm{int}}$ is the classifying space of $G_{\mathrm{int}}$, and $h^{\bullet}$ is some generalized cohomology theory, probably cobordism for bosonic systems and spin cobordism for fermionic systems\footnote{For systems with anti-unitary symmetries or non-trivial extensions of $G_{\mathrm{int}}$ by fermion parity, the precise statement is slightly modified; we will disregard this subtlety since it does not affect the conclusions.}. From axioms of generalized cohomology, and using the fact that $\mathcal{L}^* \cong \mathbb{T}^D$, one can show \cite{Else_1910} that
\begin{equation}
\label{eq:torus_classif}
    h^d(\mathcal{L}^* \times BG_{\mathrm{int}}) = \bigoplus_{k=0}^D H^k(\mathcal{L}^*, h^{d-k}(BG_{\mathrm{int}})).
\end{equation}

To interpret this equation, it is helpful to recall that, for $d-k \geq 0$, $h^{d-k}(BG_{\mathrm{int}})$ is the classification of SPTs (and invertible topological phases) with internal symmetry $G_{\mathrm{int}}$ in $d-k$ spatial dimensions. Thus, this result reduces to the classification \eqnref{eq:weak_invariants} stated in Section \ref{subsubsec:weak}. As mentioned there, we can think of the terms of \eqnref{eq:torus_classif} (for $0 < k \leq d$) as a generalization of the ``weak invariants'' that classify SPT phases with translation symmetry, associated with stacking lower-dimensional SPT phases. 

Intriguingly, however, the sum in \eqnref{eq:torus_classif} can also contain terms with $k > d$, which are not a straightforward generalization of weak invariants. In particular, one can argue that if $G_{\mathrm{int}}$ contains an anti-unitary symmetry, then $h^{-1}(BG_{\mathrm{int}}) = \mathbb{Z}_2, h^{-2}(BG_{\mathrm{int}}) = 0$; whereas, if $G$ does \emph{not} contain an anti-unitary symmetry, then
$h^{-1}(BG_{\mathrm{int}}) = 0, h^{-2}(BG_{\mathrm{int}}) = \mathbb{Z}$. To show this, one can consider the $d=0$ case of the classification \eqnref{eq:torus_classif}; here we no longer interpret $\mathcal{L}^*$ as reflecting actual phonon and phoson modes, just some manifold that parameterizes the ground state of the system. If we set $\mathcal{L}^* = S^1$, then for time-reversal invariant systems, there must be a $\mathbb{Z}_2$ topological invariant corresponding to the Berry's phase of the ground state around the circle being $0$ or $\pi$; while if we set $\mathcal{L}^*$ to be a 2-torus, then for non-time-reversal-invariant systems there be must be a $\mathbb{Z}$ topological invariant corresponding to the Chern number of the Berry connection over the torus. One can further argue that there are no new invariants arising in the $d=0$ case if $\mathcal{L}^*$ is a higher-dimensional torus, so we can conclude that $h^{r}(BG_{\mathrm{int}}) = 0$ for $r < -2$. The terms in \eqnref{eq:torus_classif} corresponding to $k > d$ reflect the theta- and Wess-Zumino terms that can appear in the elasticity theory, as we discussed in Section~\ref{sec:qc_topoterm}; see in particular Ref.~\cite{Kapustin_2001} for a microscopic discussion of the topological invariant for ground states parameterized by a space $X$ that gives rise to a Wess-Zumino term on $X$.

Finally, let us note that it is not necessary to restrict ourselves only to invertible states. We can also discuss quasicrystalline-enrichment for non-invertible topological phases; we discuss this below as part of the general theory, taking into account point-group symmetries as well.

\subsection{With point-group symmetry}
\label{sec:pgs}

We also can extend the above considerations to incorporate point-group symmetry. We first need to recall the somewhat subtle notion of point-group symmetry in a quasicrystal~\cite{Mermin_1992}. Let $R \in \mathrm{O}(d)$ be some (possibly improper) rotation, and recall from the previous sections that for a quasicrystal, there is a family of ground states labelled by the ``order parameter manifold'' $\mathcal{L}^*$. We say that a quasicrystal has point-group symmetry $R$ there is an action of $R$ on the Hilbert space that permutes these ground states among themselves. This ensures, for example, that the intensity of the peaks of the Fourier transform of the expectation value of any local observables is invariant under $R$.

Specifically, let $\ket{\Psi[\phi]}$ be the family of quasicrystalline ground states labelled by $\phi \in \mathcal{L}^*$. Let $G$ be some group, let $\rho : G \to \mathrm{O}(d)$ and $\gamma : G \to \mathcal{L}^*$ be homomorphisms, and let $U(g)$, $g \in G$ be an (anti-)unitary representation of $G$ on the Hilbert space of the system that maps operators supported near the point $\mathbf{x}$ to operators supported near the point $\rho(g) \mathbf{x}$. Then we say that the system has the symmetry $G$ if 
\begin{equation}
\label{eq:ptgroup}
U(g) \ket{\Psi[\phi]} = \ket{\Psi[\phi + \gamma(g)]}.
\end{equation}

Let us remark that, upon identifying $\mathcal{L}^*$ with $\mathbb{R}^D / \mathbb{Z}^D$, it is always possible to find a group $\widehat{G}$ and an affine-linear action of $\widehat{G}$ on $\mathbb{R}^D$, i.e.
\begin{equation}
    \mathbf{x} \mapsto R(g) \mathbf{x} + \widehat{\gamma}(g), \quad g \in \widehat{G},
\end{equation}
where $R(g) \in \mathrm{GL}(D,\mathbb{Z})$ and $\widehat{\gamma}(g) \in \mathbb{R}^D$, with the following properties: (a) $\widehat{G}$ contains all the unit translations $\mathbb{Z}^D$ as a normal subgroup, and $\widehat{G} / \mathbb{Z}^D \cong G$; (b) $\sigma(\widehat{\gamma}(g)) = \gamma(\pi(g))$ for all $g \in \widehat{G}$, where $\sigma : \mathbb{R}^D \to \mathbb{R}^D/\mathbb{Z}^D$ and $\pi : \widehat{G} \to G$ are the projection maps; and (c) the action of $R(g) \in \mathrm{GL}(D,\mathbb{Z})$ on the torus $\mathcal{L}^* = \mathbb{R}^D/\mathbb{Z}^D$ agrees with the action induced on $\mathcal{L}^*$ from the action of $\rho(g)$ on $\mathcal{L}$. In terms of the superspace picture, we can think of $\widehat{G}$ as the space group of the $D$-dimensional crystal through which we take a cut to obtain a $d$-dimensional quasicrystal. The subgroup $\mathbb{Z}^D \leq \widehat{G}$ is the translation symmetry of the $D$-dimensional crystal, and $G = \widehat{G}/\mathbb{Z}^D$ is its point group.

What we will see is that $\widehat{G}$ can be thought of as the ``effective symmetry group'' of the quasicrystal from the point view of the classification, even though there is no sense in which $\widehat{G}$ can literally be interpreted as a symmetry of the quasicrystal.
To show the classification, we note that the problem of how to classify topological families satisfying \eqnref{eq:ptgroup} is precisely the problem considered in Ref.~\cite{Thorngren_1612} (some subtleties glossed over in Ref.~\cite{Thorngren_1612} were clarified in Ref.~\cite{Else_1810}), even though the physical interpretation given in Refs.~\cite{Thorngren_1612,Else_1810} was different. Moreover, Refs.~\cite{Thorngren_1612,Else_1810} focussed on the case where the family is parameterized by a $d$-dimensional torus $\mathbb{T}^d$, or equivalently by Euclidean space $\mathbb{R}^d$. Here the family is parameterized by the space $\mathcal{L}^* \cong \mathbb{T}^D$ of dimension $D > d$. However, it was not in fact essential for any of the arguments of Refs.~\cite{Thorngren_1612,Else_1810} that the parameterizing space have dimension $d$. Therefore, by similar arguments to Refs.~\cite{Thorngren_1612,Else_1810}, we obtain

\begin{framed}
\textbf{Quasicrystalline equivalence principle}. The classification of quasicrystalline topological phases in $d$ spatial dimensions with ``effective symmetry'' $\hat{G}$ is in one-to-one correspondence with the classification of topological phases with \emph{internal} symmetry $\hat{G}$ in $d$ spatial dimensions.
\end{framed}
This is the quasicrystalline generalization of the ``crystalline equivalence principle'' of Ref.~\cite{Thorngren_1612}, and holds both for symmetry-protected and symmetry-enriched topological phases. As in the crystalline case there are some ``twists;'' for example, a unitary spatial orientation-reversing symmetry will map to an anti-unitary internal symmetry.

We emphasize that we are \emph{not} saying that quasicrystallinity is equivalent to additional spatial dimensions. Both sides of the correspondence relate to topological phases in $d$ spatial dimensions; it is the only the effective symmetry group $\widehat{G}$ that can be interpreted, if one likes, as the space group of a fictitious crystal in $D$ spatial dimensions.

Moreover, from the arguments of Refs.~\cite{Else_1810}, we can also derive a ``defect network'' picture for quasicrystalline topological phases. The defect network lives in the $D$-dimensional superspace and is required to be invariant under the symmetry $\widehat{G}$, but it is \emph{not} equivalent to a defect network for a $D$-dimensional crystalline topological phase. The difference is that in the latter, an $r$-dimensional defect carries the data of an $r$-dimensional SPT (in the invertible case, say), whereas in the former, an $r$-dimensional defect carries the data of an $[r-(D-d)]$-dimensional SPT.


\section{Conclusions}
\label{sec:cncls}

In this work, we have described a general approach for understanding many-body topological phases of matter protected by quasicrystallinity, leading ultimately to a general classification result. More concretely, we have focused on the physical implications of a particular class of quasicrystalline topological phases; namely, those protected by $\UU(1)$ charge conservation. For such phases, we have provided various interpretations and implications through the lens of topological elasticity theory. In the future, it will be desirable to explore the physical characteristics of other quasicrystalline topological phases that result from our general classification, particularly those with the ``Wess-Zumino'' type response discussed in Section~\ref{sec:qc_topoterm}; such phases are fundamentally new to quasicrystals and do not occur in crystals. Another important direction will be to search for these phases of matter in experimental platforms. As a step in this direction, it would also be useful to find concrete microscopic models realising such phases.

Here, we have focused on the bulk manifestations of quasicrystalline topological phases. It would also be interesting to consider whether the surfaces of these phases host protected gapless edge modes. For phases protected by quasicrystalline point-group symmetries, one presumably expects some kind of corner or hinge modes, as in the case of crystalline HOTIs. For non-interacting systems, this has already been verified in Refs.~\cite{varjas2019,chen2020,spurrier2020}. Meanwhile, phases such as the one depicted in Figure~\ref{fig:weak1d} (more generally, any of the phases classified by terms of \eqnref{eq:weak_invariants} with $0 < k < d$) are generalizations of crystalline weak SPT phases that host gapless edge modes provided that the surface preserves some subgroup of lattice translation symmetry. In order to extend this to quasicrystals, one would first need to develop a theory of quasicrystalline surfaces, including formulating a surface property that generalizes ``preserving a subgroup of lattice translation symmetry'' to the quasicrystalline case.

It would be interesting to generalize the quasicrystalline LSMOH result discussed in Section~\ref{sec:lsmoh} above. For crystalline systems, in cases where LSMOH forbids a trivial gapped ground state, a gapless ground state must still obey constraints such as Luttinger's theorem; such constraints were recently placed on a general footing in Ref.~\cite{Else_2007}. It might be possible to combine our results with those of Ref.~\cite{Else_2007} to obtain analogues of such constraints on gapless systems for the quasicrystalline case.

With regards to the general classification scheme discussed here, further developing the ``defect network" picture (discussed in Sec.~\ref{sec:pgs}) as well as the the ``building block" picture (developed for cSPT phases in Refs.~\cite{song2017,buildingblock}) could shed further light on the nature of quasicrystalline topological phases. Finally, it will be instructive to study non-invertible quasicrystalline phases in more detail as well, since the interplay of quasicrystallinity with topological order has yet to receive much attention. 

\begin{acknowledgements}

We thank Paul Steinhardt and Ryan Thorngren for helpful discussions. D.V.E.\ was supported by the EPiQS Initiative of the Gordon and Betty Moore Foundation, Grant No. GBMF8684. A. G. was supported by NSF CAREER Award DMR-2045181. A.P. is supported by a fellowship at the PCTS at Princeton University. S.-J.H. acknowledges support from a JQI postdoctoral fellowship and the Laboratory for Physical Sciences.
\end{acknowledgements}

\appendix


\section{Topologically equivalent Local Isomorphism classes}
\label{sec:LI}

In this Appendix, we comment on how the classification discussed in the main text relates to existing notions of equivalence classes for quasicrystals. First, it is important to understand which quasicrystals are physically indistinguishable i.e., have the same diffraction pattern and correlation functions. For periodic crystals, there is a unique arrangement of the fundamental repeating units (atoms or tiles/unit cells) that forms the ideal crystal, up to translations and rotations. In contrast, for quasicrystals there is an infinite number of ways of arranging the repeating units to form the ideal structure---the choice of orientational symmetry and the fundamental repeating units (be they atoms or tiles) does not suffice to uniquely specify a quasicrystal~\cite{socolar86}. Indeed, there exists an uncountable infinity of distinguishable arrangements of the same repeating units whose diffraction patterns (Fourier spectra) are given by the same set of reciprocal wave vectors with different Bragg peak intensities.

It is hence desirable to organise quasicrystals into equivalence classes. One such existing notion is that of \textit{local isomorphism} (LI) classes~\cite{levine86,socolar86b}, where two quasicrystals are said to be in the same LI class iff any bounded configuration of repeating units present in one appears, up to finite translations (plus global rotations and inversions), in the other with the same frequency. Quasicrystals within the same LI class cannot be distinguished by measurements made on any finite length scale. In fact, one can show that two quasicrystals are locally isomorphic if and only if they have identical diffraction patterns: the locations of the Bragg peaks \textit{and} the peak intensities of locally isomorphic quasicrystals match~\cite{levine86,socolar86b}. On the other hand, two quasicrystals whose diffraction patterns have Bragg peaks in the same locations but with differing peak intensities belong to distinct LI classes. Consequently, quasicrystals have the same free energy if they belong to the same LI class. Note that for periodic crystals, for which there exists a unique arrangement (up to translations) of the fundamental repeating units, each LI class contains a single element. 

In the atomic surface description of quasicrystals discussed in Sec.~\ref{sec:intersection}, notice that once we fix an orientation for $\Epar$, we can automatically generate an infinite number of different quasiperiodic arrangements. However, for a generic atomic surface, arrangements obtained from two cuts belong to the same LI class if and only if they can be mapped onto each other by a translation in superspace, as these shifts do not alter the diffraction pattern~\cite{numbertheory92}. For example, we can consider a translation of the two-dimensional periodic structure in Fig.~\ref{fig:tile} with respect to the origin $O$ of $\Epar$ by a vector $\mathbf{v}$. It can be shown that the resulting atomic configurations on $\Epar$ before and after the translation overlap out to arbitrary finite distances by a finite translation along $\Epar$ i.e., they belong to the same LI class and are physically indistinguishable.

Hence, LI classes subdivide quasicrystals such that elements of a given LI class are physically equivalent i.e., their diffraction patterns have Bragg peaks at identical locations and have the same intensities. As mentioned earlier, this classification is not particularly meaningful for periodic crystals where each LI class contains a single element. Nevertheless, one can still define \textit{topological} equivalence classes for periodic crystals. As discussed in Sec.~\ref{sec:crystal_elasticity}, for the case where the only symmetries are translation and $\UU(1)$ charge conservation, the integer-valued invariant $\nu$ (see Eqs.~\eqref{eq:c-topterm1}-\eqref{eq:c-topterm3}) labels the topologically inequivalent classes, where elements within a class can be smoothly deformed into each other without changing $\nu$.

\begin{figure}[t]
    \centering
    \includegraphics[width=0.8\columnwidth]{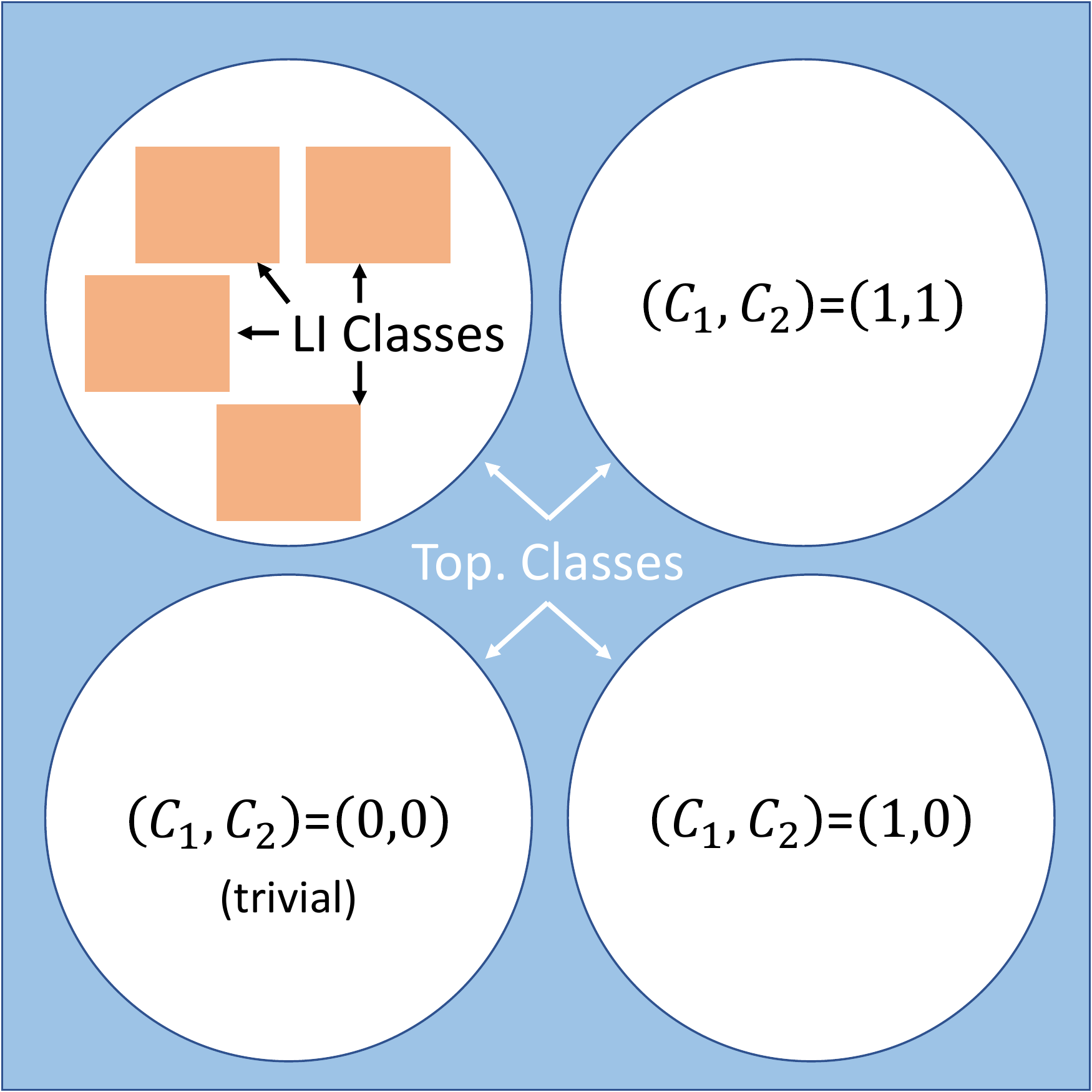}
    \caption{Topological equivalence classes for $d=1$ quasicrystals  are labelled by a pair of integers $(C_1,C_2)$ (see Secs.~\ref{sec:qc_topoterm} and~\ref{sec:interpret}). Each class consists of physically inequivalent LI classes which are equivalent topologically.}
    \label{fig:classes}
\end{figure}

This notion of \textit{topological} equivalence classes generalises to quasicrystals. As discussed earlier, there exist infinitely many arrangements of the same repeating units whose Fourier spectra have Bragg peaks in the same positions but with different peak intensities. We subdivide this space of quasicrystals into topological classes such that each class consists of LI classes that can be smoothly deformed into each other. In particular, two physically inequivalent LI classes are topologically equivalent iff one is smoothly connected to the other via deformations that leave invariant the positions of the Bragg peaks but are allowed to modify the peak intensities. Recall that deformations that also do not change the peak intensities do not change the LI class. Hence, 
our classification can be understood as further defining an equivalence relation over the space of LI classes of quasicrystals (see Fig.~\ref{fig:classes}).

\section{2d octagonal quasicrystals}
\label{sec:octagonal}

In this appendix, we show how the reciprocal lattice vectors $\mathbf{K}^I$ are obtained for a 2D octagonal quasicrystal. The idea is to obtain a decomposition $\mathbb{R}^{4} = \Epar \oplus \Eperp$  where in the 4-dimensional ``superspace,'' the physical system lives on the 2-dimensional plane $\Epar$. 

Let $a^{I}$ ($I=1,\cdots,4$) denote the basis vectors of the unit hypercubic lattice $\mathbb{Z}^{4}$. The $C_{8}$ rotation matrix in this basis is given by 
\begin{equation}
R_{8}= \left(
\begin{array}{cccc}
0 & 0 & 0 & -1 
\\
1 & 0 & 0 & 0 
\\
0 & -1 & 0 & 0 
\\
0 & 0 & -1 & 0 
\end{array}
\right).
\end{equation}
Diagonalizing $R_{8}$, we find that its eigenvalues are given by $\xi^{n} = \exp{(2 \pi i n/8)}$, where $n=1,\cdots,4$. Note that there are two conjugate pairs: $\left(\xi, \xi^{-1} \right)$ and $\left(\xi^{2}, \xi^{-2}\right)$. This implies that the superspace $\mathbb{R}^{4}$ decomposes into two orthogonal two-dimensional planes, with both planes invariant under $C_{8}$ rotation. The eigenvectors corresponding to $\xi^{n}$ ($n=1,\cdots,4$) are
\begin{equation}
v_{n} = (\xi^{n},\xi^{2n},\xi^{3n},\xi^{4n}).
\end{equation}

Although these eigenvectors have complex entries, it is possible to find a basis such that they contain only real entries. The basis vectors for one of the two-dimensional planes, which we choose to be our physical plane $\Epar$, are given by
\begin{eqnarray}
v_{1} &=& (-\frac{1}{\sqrt{2}},0,-\frac{1}{\sqrt{2}},1) ,
\\
v_{2} &=& (\frac{1}{\sqrt{2}},1, -\frac{1}{\sqrt{2}}, 0).
\end{eqnarray}
The basis vectors for the perpendicular plane $\Eperp$ are
\begin{eqnarray}
v_{3} &=& (-\frac{1}{\sqrt{2}},0,-\frac{1}{\sqrt{2}}, -1) ,
\\
v_{4} &=& (-\frac{1}{\sqrt{2}}, 1, \frac{1}{\sqrt{2}}, 0).
\end{eqnarray}
Let $\boldsymbol{M}$ be the transformation matrix between the two sets of basis vectors: $a^{I} = M\indices{^I_J} v_{J}$, where $a^{I}$ span a four-dimensional hypercubic reciprocal lattice and $v_{J}$ are orthonormal unit vectors of the four-dimensional superspace. We find that the transformation matrix is
\begin{equation}
\boldsymbol{M} = \left(
\begin{array}{cccc}
-1/\sqrt{2} & 1/\sqrt{2} & -1/\sqrt{2} & -1/\sqrt{2}
\\
0 & 1 & 0 & 1
\\
-1/\sqrt{2} & -1/\sqrt{2} & -1/\sqrt{2} & 1/\sqrt{2}
\\
1 & 0 & -1 & 0 
\end{array}
\right).
\end{equation}
The corresponding direct lattice is spanned by four vectors $a^{*I} = a^{I}/2$. By definition, the reciprocal lattice vectors are given by $ K\indices{^I_i} = 2\pi M\indices{^I_i}$ for $i=1,2$:
\begin{equation}
\boldsymbol{K} = \left(
\begin{array}{cc}
-1/\sqrt{2} & 1/\sqrt{2}
\\
0 & 1 
\\
-1/\sqrt{2} & -1/\sqrt{2}
\\
1 & 0 
\end{array}
\right).
\label{eqn:octagonal_K}
\end{equation}
The projections of the the basis vectors $\boldsymbol{a}^{*}$ of the direct lattice $\mathbb{Z}^{4}$ into the physical plane $\Epar$ are given by:
\begin{eqnarray}
\pi(a^{*}_{1}) &=& -\frac{1}{2\sqrt{2}} v_{1} + \frac{1}{2\sqrt{2}} v_{2},
\nonumber
\\
\pi(a^{*}_{2}) &=&  \frac{1}{2} v_{2}, \nonumber
\\
\pi(a^{*}_{3}) &=& -\frac{1}{2\sqrt{2}} v_{1} - \frac{1}{2\sqrt{2}} v_{2},
\nonumber
\\
\pi(a^{*}_{4}) &=& \frac{1}{2} v_{1}.
\label{eqn:v}
\end{eqnarray}
This information will be useful in constructing the tiles of the octagonal quasicrystals in a later Appendix.


\section{Density of tiles in a quasicrystalline tiling}
\label{appendix:tile_density}
In this Appendix, we will derive the formula \eqnref{eq:rhoIJ2d} for the density of a given tile in a quasicrystalline tiling generated according to the scheme of Ref.~\cite{Oguey1988}. As shown in Ref.~\cite{Oguey1988}, such a tiling can be obtained via a cut-and-project scheme. That is, one considers some $d$-dimensional hyperplane in a $D$-dimensional superspace, which can be identified with the physical space $\mathbb{R}^d$ via a linear mapping $K : \mathbb{R}^d \to \mathbb{R}^D$. Then, one 
 defines an acceptance window surrounding this hyperplane. All the $d$-dimensional facets of the $D$-dimensional hypercubic lattice (with lattice constant $2\pi$) that lie wholly within the acceptance window are projected onto the image of $K$, and the pre-image of these projections in $\mathbb{R}^d$ forms the tiles in a quasiperiodic tiling of $\mathbb{R}^d$. There are ${D \choose d}$ different facet orientations in the $D$-dimensional hypercubic lattice, labelled by unordered $d$-tuples of indices that can range over $1, \cdots, D$. The projected version of these facets then form the different kinds of tiles in the quasicrystalline tiling.

Here, what we want to prove is that in $d=2$, the density $\rho^{IJ}$ of such tiles satisfies (in some sign convention):
\begin{equation}
\label{eq:appendix_rho2d}
    \rho^{IJ} = \frac{1}{(2\pi)^2} K\indices{^I_i} K\indices{^J_j} \epsilon^{ij}.
\end{equation}
We will invoke the fact that, as shown in Ref.~\cite{Oguey1988}, the \emph{unprojected} versions of the facets contained within the acceptance window defines a closed $d$-dimensional surface in superspace, which we call the \emph{unprojected surface}.

As a warmup, let us consider the $d=1$ case, where the analogous formula to \eqnref{eq:appendix_rho2d} is:
\begin{equation}
\label{eq:appendix_rho1d}
    \rho^{I} = \frac{1}{2\pi} K\indices{^I}.
\end{equation}
Now, consider some interval of length $V$ within the 1-dimensional physical space. Then after mapping the physical space into superspace, the endpoints of this interval are separated by a displacement vector in superspace of
\begin{equation}
\label{eq:first_to_equate}
\vec{K} V,
\end{equation}
where $\vec{K}$ is the vector in superspace whose components are $K^I$.
Meanwhile, the portion of the unprojected surface that is near the mapped interval consists of a curve with endpoints, where the endpoints are separated by a displacement
\begin{equation}
\label{eq:second_to_equate}
    2\pi N^I \vec{e}_I,
\end{equation}
where $\vec{e}_I$ is the basis vector in the $I$-th coordinate direction, and $N^I$ is the number of tiles of type $I$ in the quasicrystalline tiling that are contained within the interval of length $V$. Because the unprojected surface lies within the acceptance window, whose width is independent of $V$, we conclude that \eqnref{eq:first_to_equate} and \eqnref{eq:second_to_equate} must be equal up to a correction that is $O(1)$ in $V$. Hence, by taking the $I$-th component, we find
\begin{equation}
    \frac{N^I}{L} = \frac{1}{2\pi} K^I + O\left(\frac{1}{V}\right),
\end{equation}
and taking the limit $V \to \infty$ gives \eqnref{eq:appendix_rho1d}.

The $d=2$ version of the argument proceeds similarly, except that the quantity that we need to equate between the physical plane and the unprojected surface is the ``net directed area,'' as measured by a bivector (i.e. a rank-2 element of the exterior algebra of $\mathbb{R}^D$). Thus, we require that
\begin{equation}
    (2\pi)^2 N^{IJ} \vec{e}_I \wedge \vec{e}_J = \epsilon^{ij} V \vec{K}_i \wedge \vec{K}_j + O(1),
\end{equation}
where $\vec{K}_i$ is the vector in $\mathbb{R}^D$ given by taking the $i$-th column of the matrix $K$.
Taking components of this equation gives \eqnref{eq:appendix_rho2d}.



\section{Ammann-Beenker tiling}
\label{sec:abtiling}

A 2d octagonal quasicrystal can be constructed by the Ammann-Beenker tilings. In an Ammann-Beenker tiling, there are two kinds of tiles: a rhombus (with the small angle equal to $\pi/4$), and a square. Here, we briefly explain how to obtain these tiles by the projection method. We are going to follow Ref.~[\onlinecite{Oguey1988}] closely. 

\begin{figure}[t]
\includegraphics[width=0.8\columnwidth]{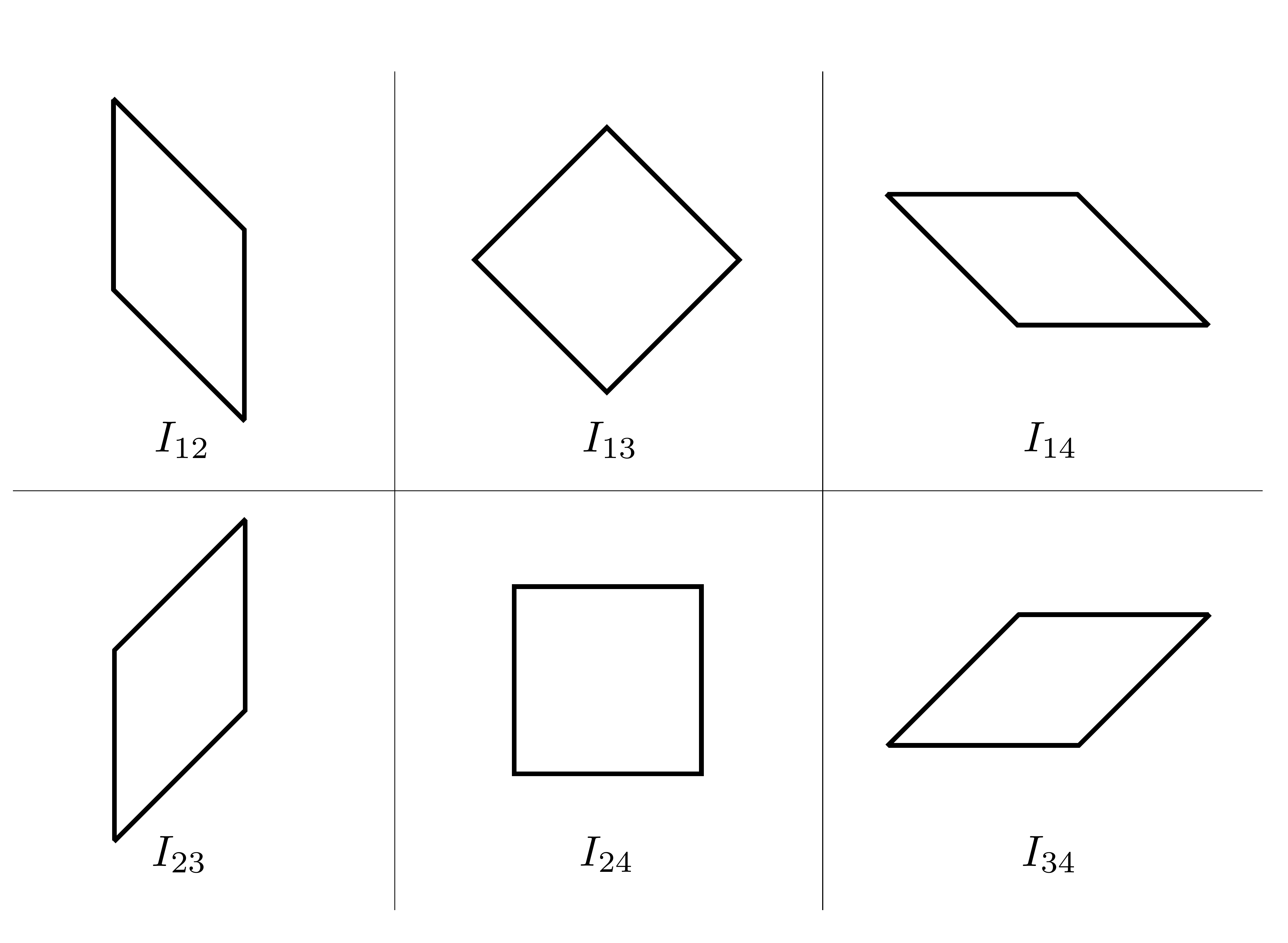}
\caption{Prototiles of the Ammann-Beenker tiling and their indices. 
}
\label{fig:protile_AB}
\end{figure}

The tiles are simply given by the projection of the 2-facets of a $D=4$ hypercubic unit cell onto the physical plane. Let $M_{2} = \{I=\{i_{1},i_{2}\} \subset \{1,2,3,4\}\}$, be the set of subsets of $\{1,2,3,4\}$. The 2-facets of the unit hypercube, indexed by $M_{2}$, are defined as follows:
\begin{equation}
\gamma_{I} = \{ \sum_{i \in I} \lambda_{i}a^{*}_{i} | \lambda_{i} \in [0,1] \} ,
\end{equation}
for all $I \in M_{2}$.
Now, we use Eq.~\eqref{eqn:v} to obtain the projections of the facets $\gamma$ onto the physical plane:
\begin{equation}
D_{I} = \pi(\gamma_{I}) = \{ \sum_{i \in I} \lambda_{i} v_{i}^{P} | \lambda_{i} \in [0,1]\},
\end{equation}
where we have defined $v_{i}^{P} := \pi(a^{*}_{I})$. Since a tile in 2d requires $2$ coordinates, there are ${4 \choose 2} = 6$ prototiles. For example, a rhombus indexed by $I_{12} := \{ 1,2 \}$ is 
\begin{equation}
D_{I_{12}} = \lambda_{1} v^{P}_{1} + \lambda_{2} v^{P}_{2}.
\end{equation}
A square indexed by $I_{24} := \{2,4\}$ is 
\begin{equation}
D_{I_{24}} = \lambda_{2} v^{P}_{2} + \lambda_{4} v^{P}_{4}.
\end{equation}
Fig.~\ref{fig:protile_AB} shows the prototiles of the Ammann-Beenker tiling along with their index. Note that, if we include the $C_{8}$ rotation in the definition of the congruence classes, there are only two prototiles: the rhombus and the square. The Ammann-Beenker tiling shown in Fig.~\ref{fig:weakdots} can be constructed by using the prototiles shown in Fig.~\ref{fig:protile_AB}.

\section{Deriving the mobility constraint of dislocations in a 2D quasicrystal}
\label{appendix:mobility}
In this Appendix, we give a more precise argument for the constraint \eqnref{eq:mobilityconst} on the motion of dislocations in a 2D quasicrystal.

The $\theta$ field for a dislocation moving uniformly in space, with its core at position $\mathbf{a}(t)$ (we assume that $\mathbf{a}(0) = \mathbf{0}$) can be written as
\begin{equation}
\theta^I(\mathbf{x},t) = \Theta^I(\mathbf{x},t) + \phi^I(\mathbf{x},t),
\end{equation}
where we have defined $\Theta^I(\mathbf{x}) = K^I_i x^i$, and $\phi$ satisfies
\begin{equation}
\label{eq:motion_property}
\phi^I(\mathbf{x},t) = \phi^I(\mathbf{x} - \mathbf{a}(t),0).
\end{equation}
Let $\Sigma$ be a loop in 2D space encircling the dislocation at time $t=0$, and let $\Sigma(t) = \Sigma + \mathbf{a}(t)$. Now define $\Sigma_+[0,T]$ to be the 2D worldsheet in 3D space-time swept out by $\Sigma(t)$ between time $t=0$ and time $t=T$. We can evaluate the amount of charge that must have been created at the dislocation core during the motion from time $t=0$ to time $t=T$ by evaluating the integral
\begin{equation}
\Delta Q = \int_{\Sigma_+[0,\tau]} (*J).
\end{equation}
where $J$ is the 1-form in 3D space-time constructed by using the metric to lower the index of the current 3-vector $J^\mu$, and $*$ is the Hodge star operator. Here, and in what follows, we use the abstract notation for differential forms to simplify the derivation.
From \eqnref{eqn:current} we have that
\begin{align}
*J &= \frac{1}{8\pi^2} C_{IJ} d\theta^I \wedge d\theta^J \\
   &= \frac{1}{8\pi^2} (C_{IJ} K^I \wedge K^J + 2C_{IJ} K^I \wedge d\phi^I + d\phi^I \wedge d\phi^J).
\end{align}
The integral of the first term gives zero, while the integral of the third term can also be shown to be zero by first showing that it is invariant under small deformations of $\phi^I$ (provided that \eqnref{eq:motion_property} remains satisfied) and then deforming $\phi^I$ to a reference configuration for a particular winding number sector and evaluating the integral.

This leaves only the integral of the second term, and hence we find:
\begin{align}
4\pi^2 \Delta Q &= \int_{\Sigma_+[0,T]} C_{IJ} K^I \wedge d\phi^J \\
         &= \int_{\Sigma_+[0,T]} C_{IJ} d\Theta^I \wedge d\phi^J \\
         &= \int_{\Sigma_+[0,T]} C_{IJ} d(\Theta^I d\phi^J) \\
         &= \int_{\partial \Sigma_+[0,\tau]} C_{IJ} \Theta^I d\phi^J \label{eq:used_stokes} \\
         &= C_{IJ} \biggl[\int_{\Sigma(T)}  K\indices{^I_i} x^i [\nabla \phi^J(\mathbf{x}, T)] \cdot d\mathbf{x} \nonumber \\
         &\quad \quad \quad \quad - \int_{\Sigma} K\indices{^I_i} x^i [\nabla \phi^J(\mathbf{x},0)] \cdot d\mathbf{x} \biggr] \label{eq:hello} \\
         &= C_{IJ} K\indices{^I_i} a^i(T) \int_{\Sigma} [\nabla \phi^J(\mathbf{x},0)] \cdot d\mathbf{x} \label{eq:goodbye} \\
        &=  C_{IJ} K\indices{^I_i} a^i(T) b^J,
\end{align}
where in \eqnref{eq:used_stokes} we used Stokes' theorem, and to go from \eqnref{eq:hello} to \eqnref{eq:goodbye} we invoked \eqnref{eq:motion_property} and made the change of variables $\mathbf{x} \to \mathbf{x} + \mathbf{a}(T)$ in the first term. Here $b^J$ is the ``Burgers vector'' of the dislocation defined by \eqnref{eqn:dislocation}. If we demand that $\Delta Q = 0$ and take the limit of infinitesimal $T$, we recover the constraint \eqnref{eq:mobilityconst}.


\bibliography{ref-manual,ref-autobib}

\end{document}